\documentclass[journal,12pt, draftclsnofoot, onecolumn,romanappendices]{IEEEtran}
\usepackage{cite}
\usepackage{graphicx}  
\usepackage{subcaption}
\usepackage{caption}
\DeclareCaptionFont{tiny}{\tiny}
\usepackage{epstopdf}
\usepackage{amssymb,times}
\usepackage{centernot}

\usepackage{algorithm}
\usepackage{algorithmic}

\usepackage{multirow}
\usepackage{psfrag}

\usepackage[cmex10]{amsmath}
\usepackage{amssymb}
\usepackage{color}
\usepackage{amsthm}

\def\diag{\mathrm{diag}}

\newtheorem{theorem}{Theorem}

\newtheorem{corollary}{Corollary}

\newtheorem{assumption}{Assumption}

\usepackage[dvipsnames,svgnames]{xcolor}
\usepackage[textwidth=30mm]{todonotes}

\newcommand{\be}{\begin{equation}}
\newcommand{\ee}{\end{equation}}
\newcommand{\ba}{\begin{array}}
\newcommand{\ea}{\end{array}}
\newcommand{\bea}{\begin{eqnarray}}
\newcommand{\eea}{\end{eqnarray}}
\newcommand{\herm}{^{\mbox{\scriptsize H}}}

\newcommand{\tran}{^{\mbox{\scriptsize T}}}

\newcommand{\vbar}{\raisebox{.17ex}{\rule{.04em}{1.35ex}}}
\newcommand{\vbarind}{\raisebox{.01ex}{\rule{.04em}{1.1ex}}}
\newcommand{\R}{\ifmmode {\rm I}\hspace{-.2em}{\rm R} \else ${\rm I}\hspace{-.2em}{\rm R}$ \fi}
\newcommand{\T}{\ifmmode {\rm I}\hspace{-.2em}{\rm T} \else ${\rm I}\hspace{-.2em}{\rm T}$ \fi}
\newcommand{\N}{\ifmmode {\rm I}\hspace{-.2em}{\rm N} \else \mbox{${\rm I}\hspace{-.2em}{\rm N}$} \fi}
\newcommand{\B}{\ifmmode {\rm I}\hspace{-.2em}{\rm B} \else \mbox{${\rm I}\hspace{-.2em}{\rm B}$} \fi}
\newcommand{\Hil}{\ifmmode {\rm I}\hspace{-.2em}{\rm H} \else \mbox{${\rm I}\hspace{-.2em}{\rm H}$} \fi}
\newcommand{\C}{\ifmmode \hspace{.2em}\vbar\hspace{-.31em}{\rm C} \else \mbox{$\hspace{.2em}\vbar\hspace{-.31em}{\rm C}$} \fi}
\newcommand{\Cind}{\ifmmode \hspace{.2em}\vbarind\hspace{-.25em}{\rm C} \else \mbox{$\hspace{.2em}\vbarind\hspace{-.25em}{\rm C}$} \fi}
\newcommand{\Q}{\ifmmode \hspace{.2em}\vbar\hspace{-.31em}{\rm Q} \else \mbox{$\hspace{.2em}\vbar\hspace{-.31em}{\rm Q}$} \fi}
\newcommand{\Z}{\ifmmode {\rm Z}\hspace{-.28em}{\rm Z} \else ${\rm Z}\hspace{-.28em}{\rm Z}$ \fi}

\renewcommand{\vec}[1]{\bf{#1}}     

\begin{document}


\title{Decentralizing Multicell Beamforming via Deterministic Equivalents
	}

\author{Hossein Asgharimoghaddam, \emph{Student Member, IEEE} Antti T\"olli, \emph{Senior Member, IEEE},  Luca Sanguinetti, \emph{Senior Member, IEEE}, Merouane Debbah, \emph{Fellow, IEEE}
\thanks{ \newline \indent H. Asgharimoghaddam and A. T\"olli are with the Centre for Wireless Communications, University of Oulu, Oulu, Finland (Firstname.Lastname@oulu.fi). L.~Sanguinetti is with the University of Pisa, Dipartimento di Ingegneria dell'Informazione, Italy (luca.sanguinetti@unipi.it) and with the Large Systems and Networks Group (LANEAS), CentraleSup\'elec, Universit\'e Paris-Saclay, 3 rue Joliot-Curie,  91192 Gif-sur-Yvette, France. 
M. Debbah is with the Mathematical and Algorithmic Sciences Lab, Huawei Technologies Co. Ltd., France (merouane.debbah@huawei.com).
\newline\indent 
This work was supported in part by the Academy of Finland 6Genesis Flagship (grant No. 318927),  the Academy of Finland funding No. 279101 and FP7 project ICT-317669 METIS.  L.~Sanguinetti and M. Debbah have been supported by the ERC Starting Grant 305123 MORE. 
}
}

\maketitle
\vspace{-2cm}
\begin{abstract}	
This paper focuses on developing a decentralized framework for coordinated minimum power beamforming wherein $L$ base stations (BSs), each equipped with $N$ antennas, serve $K$ single-antenna users with specific rate constraints. This is realized by considering user specific intercell interference (ICI) strength as the principal coupling parameter among BSs. First, explicit deterministic expressions for transmit powers are derived for spatially correlated channels in the asymptotic regime in which $N$ and $K$ grow large with a non-trivial ratio $K/N$. These asymptotic expressions are then used to compute approximations of the optimal ICI values that depend only on the channel statistics. By relying on the approximate ICI values as coordination parameters, a distributed non-iterative coordination algorithm, suitable for large networks with limited backhaul, is proposed. A heuristic algorithm is also proposed relaxing coordination requirements even further as it only needs pathloss values for non-local channels. The proposed algorithms satisfy the target rates for all users even when $N$ and $K$ are relatively small.
Finally, the potential benefits of grouping users with similar statistics are investigated to further reduce the overhead and computational effort of the proposed solutions. 
Simulation results show that the proposed methods yield near-optimal performance.
\end{abstract}


\IEEEpeerreviewmaketitle

\section{Introduction}
High spatial utilization is a promising approach to meet the significant spectral efficiency enhancements required for 5G cellular networks. In general, this achieved by using a large number of antennas $N$ at the base stations (BSs) to serve a large number of user equipments (UEs) $K$ on the same frequency-time resources. The need for serving such a large number of UEs in multicellular environments pronounces the importance of proper precoder design that takes into account the intercell coordination and subsequent challenges in such large networks.
In the context of massive multiple-input-multiple-output (MIMO)~\cite{marzlaryngo16,BjoHoydSang17} under the assumption of i.i.d.~Rayleigh fading channels (i.e., no spatial correlation), as $N\to\infty$ with $K$ fixed, non-cooperative precoding schemes such as maximum ratio transmission~\cite{marzetta2010noncooperative}, single-cell~\cite{hoydis2013massive,Krishnan2014a} and multicell\cite{Ngo2012b,EmilEURASIP17} minimum mean squared error (MMSE) schemes
can entirely eliminate the multicell interference and the performance of is only limited by pilot contamination. As shown recently in~\cite{BjornsonHS17}, even the pilot contamination is not a fundamental asymptotic limitation when a
small amount of spatial channel correlation or large-scale fading variations over the array is considered.
Despite all this, when $N$ is not relatively large compared to $K$, cooperation among cells provides additional benefits in mitigating intercell interference (ICI). Coordinated multicell resource allocation is generally formulated as an optimization problem in which the desired network utility is maximized subject to some constraints.
In this work, we consider a coordinated multicell multiuser MIMO
system in which $L$ BSs, each equipped with $N$ antennas, jointly minimize the transmission power required to satisfy target rates for $K$ single-antenna UEs. 
{We recognize that the ICI coordination in a dense network with a large number of UEs and antennas is very challenging, due to the practical limitations of backhaul links. Hence, we are particularly interested in developing a semi-static coordination scheme that allows the cooperating BSs to locally obtain near-optimal precoders with a minimal information exchange.}

\subsection{Prior Work}
Coordinated multicell minimum power beamforming has been largely investigated in the literature~\cite{SOCPshamai,Dahrouj-Yu-10,Tolli-Pennanen-Komulainen-TWC10, Pennanen-Tolli-Latva-aho-spl-11, ChaoADMM2012, Pennanen-Tolli-Latva-aho-TSP-13} and has recently received renewed interest in the context of green multicellular networks~\cite{GreenNet}.
The optimal solution to this optimization problem
can be computed by means of second-order cone programming (SOCP)~\cite{SOCPshamai} or exploiting uplink-downlink duality~\cite{ Dahrouj-Yu-10}. However, this requires full channel state information (CSI) at all BSs, meaning that the locally measured instantaneous CSI needs to be exchanged among BSs. To avoid the exchange of CSI among BSs, several different decentralized solutions have been proposed in the literature~\cite{Tolli-Pennanen-Komulainen-TWC10, Pennanen-Tolli-Latva-aho-spl-11, ChaoADMM2012, Pennanen-Tolli-Latva-aho-TSP-13}. The underlying idea of all these methods is to reformulate the optimization problem such that the BSs are only coupled by real-valued ICI terms. In this way, the centralized problem can be decoupled by primal or dual decomposition approaches leading to a distributed algorithm, which needs the ICI values to be continuously exchanged among BSs (to follow the changes in the fading process). Despite the remarkable reduction in information exchange, when the system dimensions grow large (as envisioned in 5G networks) and consequently the amount of information to be exchanged increases, the limited capacity and high latency of backhaul links in practical networks may become a bottleneck.

A possible way out of these issues is to rely on the asymptotic analysis in which  $N$ and $K$ grow large with a non-trivial ratio $K/N$. In these circumstances, tools from random matrix theory allow to derive explicit expressions for (most) performance metrics such that they only depend on the channel statistics~\cite{RMT}.
 The asymptotic analysis for the closely related problem, i.e., regularized zero-forcing precoding, is presented in~\cite{wagner2012large,zhang2013large}, and the power minimization problem in conjunction with sum-rate maximization is considered in~\cite{SHeY2015SumRtToPWR}.
    In the course of developing large system analysis for power minimization problem subject to UEs' rate constraints, one can begin with the Lagrangian duality formulation developed in~\cite{ Dahrouj-Yu-10} where the optimal power assignments are presented in terms of channel entries. In particular, the results of large system analysis can be utilized to compute deterministic equivalents for the optimal powers.    
The authors in~\cite{sanguinetti2014optimal,LucaSanguinetti,zakhour2013min, lakshminarayana2015coordinated, LucaCouilletDebbahJournal2015} perform such analysis under i.i.d. Rayleigh fading channels in single-cell~\cite{sanguinetti2014optimal,LucaSanguinetti} and multicell~\cite{zakhour2013min, lakshminarayana2015coordinated, LucaCouilletDebbahJournal2015} settings. The impact of spatial correlation on the asymptotic power assignment is studied in~\cite{Huang2012correlated} for a single-cell scenario with UEs experiencing identical correlation matrices. The deterministic equivalents are found to depend only on the long-term channel statistics and on the UEs' target rates. This enables the cross-cell coordination based on slow varying channel statistics and also provides insights into the structure of the optimal solution as a function of underlying statistical properties.
However, the major drawback in using the asymptotic power expressions in practical networks of finite size (with a finite number of antennas $N$) is that the rate constraints are not met since those can be guaranteed only asymptotically.
\subsection{Contributions}
{The main contribution of this paper is to introduce two novel semi-static coordination algorithms that allow BSs to obtain near-optimal QoS-guaranteed precoders locally, and subject to relaxed coordination requirements. This is realized by
		reformulating the optimization problem such that the BSs are only coupled by ICI values~\cite{Tolli-Pennanen-Komulainen-TWC10}. Then,        
		by utilizing the Lagrangian duality analysis in~\cite{Dahrouj-Yu-10} and techniques of random matrix theory~\cite{RMT}, we derive deterministic equivalents for the optimal ICIs in terms of channel statistics. They are derived under a generic spatially correlated channel model. Such an analysis is instrumental to develop two distributed algorithms.}     
	\begin{itemize}     
		\item {Algorithm~\ref{alg:ICI_approx} incorporates the deterministic ICIs as approximations for the coordination messages in finite networks. This allows the BSs to obtain the precoders locally by exploring the local CSI and exchanges of slow varying channel statistics over the backhaul links. }
		\item {
		Algorithm~\ref{alg:ICI_approx heuristic} includes a heuristic simplification in the calculation of approximate ICIs. This allows an alternative backhaul signaling that reduces the backhaul exchange rate requirement by a factor of almost $2/N^2$ compared to Algorithm~1. The performance loss of both algorithms is shown to be small, with respect to the optimal solution, via numerical results. }   
	\end{itemize}

{The large system analysis is also developed in a special scenario where UEs are assumed to be grouped on the basis of their statistical properties as in~\cite{UserPartionAdhikary,GroupKim2015}. 
This allows to derive the approximate ICIs in concise form, thereby revealing the structure of the coordination messages; that is, the optimal ICI values in terms of underlying channel statistics.}
	 {The analysis ultimately reveals the potential benefits of UE-grouping to further reduce the overhead and computational effort of the proposed decentralized solutions.}

Parts of this paper have been published in the conference
publications~\cite{Asgharimoghaddam-Tolli-Rajatheva-ICC2014,asgharimoghaddam2014decentralized,HAsghariTolliLucaDebbah2015CAMSAP}. The decentralized solution relying on deterministic ICI values is investigated in an i.i.d Rayleigh fading model in~\cite{Asgharimoghaddam-Tolli-Rajatheva-ICC2014} and in the correlated scenario in~\cite{asgharimoghaddam2014decentralized, HAsghariTolliLucaDebbah2015CAMSAP}. Specifically, the large system analysis is sketched in~\cite{asgharimoghaddam2014decentralized, HAsghariTolliLucaDebbah2015CAMSAP} while
the precise proofs along with derivation details are presented in the current work.
In addition, the analysis is extended to the case where UEs are grouped on the basis of their statistical properties.
In the numerical analysis, the exponential correlation model, utilized in the conference counterparts, is extended to a more general case where UEs experience various angle of arrivals and angular spreads. 
A numerical study for the UE-grouping scenario~is also provided.

The remainder of this work is organized as follows.\footnote{The following notations are used throughout the manuscript. All boldface letters indicate
	vectors (lower case) or matrices (upper case). Superscripts
	$(\cdot)\tran$, $(\cdot)\herm$, $(\cdot)^{-1}$, $(\cdot)^{1/2}$
	stand for transpose, Hermitian transpose, matrix inversion and
	positive semidefinite square root, respectively. We use $\mathbb{C}^{m
		\times n}$ and $\mathbb{R}^{m
		\times n}$ to denote the set of $m \times n$ complex and real valued matrices, respectively.
	Furthermore, $\diag( \cdots)$ denotes the diagonal matrix with
	elements $(\cdots)$ on the main diagonal. 
	The sets are indicated by
	calligraphic letters and $|\mathcal{A}|$ denotes the cardinality
	of the set $\mathcal{A}$. 
		${\mathrm{Tr}}({\bf A})$ denotes the trace of $\bf A$, and $\|\cdot\|$ represents the Euclidean norm. Finally, $[.]_{i,j}$ denotes the $(i,j)^{\text{th}}$ element of the matrix and $\mathcal{A}\backslash k
	$ excludes the index $k$ from the set.} In
Section~\ref{sec:System model}, the network
model and problem formulation are presented.
Section~\ref{sec: large system analysis} deals with the large system analysis of the optimal power allocations. Section~\ref{sec:distributedOpt} makes use of the asymptotic analysis to derive two distributed solutions with different coordination overheads. In Section~\ref{sec:grouped UEs}, the analysis is extended to a network in which the UE population is partitioned in groups 
on the basis of statistical properties as in~\cite{UserPartionAdhikary,GroupKim2015}.
Section~\ref{sec:Simulation Results} describes the
simulation environment and illustrates numerical results. Conclusions are drawn in Section~\ref{sec:Conclusion} while all the proofs are presented in the Appendices.

\section{{System Model And Problem Formulation}}
\label{sec:System model}
Consider the downlink of a multicell multiuser MIMO system composed of $L$ cells where each BS has $N$ antennas. A total number of $K$ single-antenna UEs is dropped in the coverage area. We assume that each UE is assigned to a single BS while being interfered by the other BSs. We denote the set of UEs served by BS $b$ as $\mathcal U_{b}$ and the index of the BS associated to UE $k$ as $b_{k}$. The set of all UEs is represented by $\mathcal U$ whereas $\mathcal B$ collects all BS indexes. { Under this convention and assuming narrow-band transmission, we define ${\bf h}_{b,k} \in \mathbb{C}^{N}$ as the channel from BS $b$ to UE $k$ and ${\bf w}_{k} \in \mathbb {C}^{N}$ as the precoding vector of UE $k$ at the intended BS. Then, the received signal can be written as}
\begin{align}
y_{k}  =  {\bf h}_{b_{k},k}\herm{\bf w}_{k}s_{k} \!+\!\!\!\! \!\!\sum\limits_{i\in{\mathcal U}_{b_{k}}\setminus k}\!\! {\bf h}_{b_k,k}\herm{\bf w}_{i}s_{i} +\!\!\! \!\!\sum\limits_{b\in{\mathcal B}\setminus b_{k}} \!\sum\limits_{i\in {\mathcal U}_{b}}{\bf h}_{b,k}\herm{\bf w}_{i}s_{i}  + n_{k}
\end{align}
where the first term is the desired received signal whereas the second and third ones represent intra-cell and inter-cell interference terms, respectively. The zero mean, unit variance data symbol intended to UE $k$ is denoted by  $s_{k}$, and is assumed to be independent across UEs. 
 Denoting the receiver noise by
$n_{k}\sim \mathcal {CN}(0,\sigma^{2})$ and treating interference as noise, the SINR attained at UE $k$ is given by
\begin{equation}\label{eq:SINR dl}
 {\Gamma}_{k}=\frac{\left|{\bf h}_{b_k,k}\herm{\bf w}_{k}\right|^{2}}{\sum\limits_{i\in {\mathcal U}_{b_{k}}\setminus k} \left|{\bf h}_{b_k,k}\herm{\bf w}_{i}\right|^{2} + \sum\limits_{b\in\mathcal{B}\backslash b_k, j\in \mathcal{U}_{b}} \left|{\bf h}_{b,k}\herm{\bf w}_{j}\right|^{2} +  \sigma^{2}}. 
\end{equation}
\subsection{Coordinated Beamforming}
{In a coordinated network, the BSs design precoders jointly to satisfy a given set of SINRs for all UEs while minimizing the total transmit power.
In order to reflect different power budgets at BSs, we consider the problem
of minimizing the weighted total transmit power with the
transmit power at BS $b$ weighted by a factor~$\mu_{b}$ as proposed in~\cite{Dahrouj-Yu-10}. This yields
\begin{equation}\label{eq:prim problemSimple}
\min_{\{{\bf w}_{k}\}} \quad \sum_{b{\in}\mathcal{B}}\sum_{k{\in}\mathcal{U}_{b}} \mu_b \|{\vec w}_{k}\|^2\quad
{\text{s.t.}}\quad {\Gamma}_k\ge \gamma_{k}, \;\forall k \in \mathcal{U}
\end{equation}    
where $\gamma_{k}$ denotes the UE's target SINR obtained from the corresponding target rate. The SINR target constraints in~\eqref{eq:prim problemSimple} may appear to be non-convex at a first glance. However, non-convex constraints of this type can be transformed into a second-order cone constraint~\cite{SOCPshamai}, which enables methods for solving~\eqref{eq:prim problemSimple} via convex optimization.
Denoting the ICI term from BS $b$ to UE $k$ as $\epsilon_{b,k}$, the optimization problem in~\eqref{eq:prim problemSimple} can be equivalently reformulated as~\cite{Tolli-Pennanen-Komulainen-TWC10}
\begin{subequations}
\label{Opt_problem}
	\begin{align}
 &\underset{{\vec w}_{k},\epsilon_{b,k}}{\min}
 \quad \sum_{b{\in}\mathcal{B}}\sum_{k{\in}\mathcal{U}_{b}} \mu_b \|{\vec w}_{k}\|^2 \\
 &\,\,\,\,\text{s.t.}
   \frac{\left|{\bf h}_{b_k,k}\herm{\bf w}_{k}\right|^{2}}{\sum\limits_{i\in {\mathcal U}_{b_{k}}\setminus k} \left|{\bf h}_{b_k,k}\herm{\bf w}_{i}\right|^{2} + \sum\limits_{b\in{\mathcal B}\setminus b_{k}} \epsilon_{b,k} +  \sigma^{2}} \geq {\gamma_{k}}, \, \forall k \in \mathcal{U}_{b} , \forall b \\
& \quad  \sum_{j \in \mathcal{U}_{b}}|{\vec h}_{b,k}\herm{\vec w}_{j}|^2 \leq \epsilon_{b,k}, \; \forall k\not\in \mathcal{U}_{b}, \forall b. \label{Opt_problem_ICICons}
	\end{align}
\end{subequations}
As shown in~\cite{Tolli-Pennanen-Komulainen-TWC10}, \eqref{eq:prim problemSimple} and \eqref{Opt_problem} are equivalent at the optimal solution where the ICI constraints in~(\ref{Opt_problem}c) are satisfied with equality. 
The problem formulation in~\eqref{Opt_problem} recognizes the ICI constraints as the principal coupling parameters among BSs, which enforces cross-cell coordination.
 Therefore, one may need to solve the problem either centrally~\cite{SOCPshamai, BjornsonMinPower2014,Dahrouj-Yu-10}, which requires full CSI at all BSs, or distributively, which needs the ICI values to be continuously exchanged among BSs~\cite{ Tolli-Pennanen-Komulainen-TWC10, Pennanen-Tolli-Latva-aho-spl-11, ChaoADMM2012, Pennanen-Tolli-Latva-aho-TSP-13}. This is hard in practice to achieve due to the practical limitations of backhaul~links.}
\subsection{Decentralized Solution Via Deterministic Equivalents}
\label{sec:distibuted Sol intro}
{We observe that using any fixed ICI term in~\eqref{Opt_problem} decouples the problems at BSs. 	
	This leads to a suboptimal solution  that, however, satisfies the SINR constraints (if feasible) subject to a higher total transmit power as compared to the optimal solution. 	Provided a set of good approximations for the optimal ICI terms $\{\epsilon_{b,k}\}$, the individual problems at the BSs can be decoupled subject to small performance loss. { In order to derive such approximations, we need to formulate the {Lagrange dual problem} of~\eqref{Opt_problem} to unveil the structure of the optimal beamfomers, and thus, express the ICIs as a function of channel entries. 	
	The authors in~\cite{Dahrouj-Yu-10} show that 
	upon existence, the unique solution\footnote{{The arguments ${{\bf w}_j,\,\forall j}$ of the problem in~\eqref{eq:prim problemSimple} are defined up to a phase scaling, i.e., if ${\bf w}_{j}$ is optimal, then ${\bf w}_{j}e^{i\phi_j}$ is also optimal~\cite{SOCPshamai} where $\phi_j$ is an arbitrary phase rotation for UE $j$. The uniqueness of the solution is declared as we can
			restrict ourselves to precoders ${{\bf w}_j,\,\forall j}$ such that the terms $\{{\bf h}_{b,k}^{\text{H}} {\bf w}_{j}\}$ in~\eqref{eq:SINR dl} have non-negative real part and a zero imaginary part.}}
	to the problem in~\eqref{eq:prim problemSimple} can be obtained by using {Lagrangian duality} in convex optimization.
		We prove that the Lagrange dual problem of~\eqref{Opt_problem} is the same as that of~\eqref{eq:prim problemSimple}, and thus, we can
	utilize the Lagrangian duality analysis in~\cite{Dahrouj-Yu-10}. To keep the flow of the work uninterrupted, the details of Lagrangian duality analysis are presented in Appendix~\ref{sec:duality analysis}. As a result of this analysis, the ICI from BS $b$ to UE $k$ can be expressed as}
	\begin{align}\label{3.1}
	\epsilon_{b,k} = \sum_{j\in \mathcal U_{b}}\left|{\bf h}_{b,k}^H{\bf w}_{j}\right|^{2} = \sum_{j\in \mathcal U_{b}}\delta_{j}\left|{\bf h}_{b,k}^H{\bf v}_{j}\right|^{2}
	\end{align}
	where, $\{{\vec v}_{k}\}$ denotes a set of  minimum mean square error (MMSE) receivers, and $\{\delta_k\}$ are scaling factors relating the beamforming vectors to the MMSE receivers as ${{\vec w}}_{k}=\sqrt{{{\delta_{k}}}/{N}} {\vec v}_{k}$. In particular, we have $
	\mathbf{v}_{k}=(\sum_{j\in \mathcal U\setminus k}{\lambda_{j}}\mathbf{  h}_{b_{k},j}\mathbf{  h}_{b_{k},j}\herm + \mu_{b_k}N\mathbf{I}_{N})^{-1}\mathbf{  h}_{b_k,k}$ with $\{\lambda_j\}$ being the Lagrange dual variables associated with SINR constraints.
	 The optimal Lagrangian multipliers gathered in $\boldsymbol{\lambda}^*= [\lambda_{1}^*, \ldots, \lambda_{K}^*]\tran $ are obtained as the unique fixed point solution of
	\begin{align}
	\label{eq:lambda itr}
	\lambda_{k} = \frac{\gamma_{k}}{ {\mathbf{h}_{b_k,k}\herm \left(\sum\limits_{j\in \mathcal U\setminus k}\lambda_{j}\mathbf{  h}_{b_{k},j}\mathbf{  h}_{b_{k},j}\herm + \mu_{b_k}N\mathbf{I}_{N}\right)^{-1}\!\!\!\!\!\!\mathbf{h}_{b_k,k}} } \,\,\forall k \in \mathcal{U}.
	\end{align}
	The scaling factors $\{\delta_{k}\}$ can be obtained as the unique solution of the set of equations such that the SINR
	constraints in~\eqref{eq:prim problemSimple} are all satisfied, i.e., $\boldsymbol{\delta} =  \mathbf{G}^{-1} \mathbf{1}_{K}\sigma^2$, where $\boldsymbol{\delta} = [\delta_{1}, \ldots, \delta_{K}]\tran$, and the $(i,k)^{\text{th}}$ element of the so-called coupling matrix $\mathbf{G} \in \mathbb C^{K\times K}$ is~\cite{ Dahrouj-Yu-10}
	\begin{align}\label{eq:G_matrix}
	\left[\mathbf{G}\right]_{k,i}= \begin{cases}
	\frac{1}{\gamma_{k}}{|\mathbf{h}_{b_k,k}\herm \mathbf{v}_{k}|^2}& \text{for} \,\,\, i=k \\
	-{|\mathbf{h}_{b_{i},k}\herm \mathbf{v}_{i}|^2}& \text{for} \,\,\, i\ne k.
	\end{cases}
	\end{align}}
The optimal ICIs in~\eqref{3.1} are expressed in terms of channel entries via  parameters $\{\lambda_k\}$,  $\{\delta_k\}$ and $\{[{\bf G}]_{i,j}\}$.
This allows us to utilize  techniques for deterministic equivalents~\cite{RMT}, as detailed in Section~\ref{sec: large system analysis}, to characterize the behavior of these parameters in terms of underlying channel statistics, and thus, propose proper approximations for the optimal ICIs.
To this end, we first need to introduce a statistical model for channel vectors. 
\subsection{Channel Model}
\label{sec:Ch model}
{The channel from BS $b$ to UE $k$ is modeled as ${\bf h}_{b,k} = {\bf \Theta}_{b,k}^{1/2}{\bf z}_{b,k}$ where 
${\bf z}_{b,k}\in  \mathbb {C}^{N}$ represents small-scale fading and has i.i.d, zero-mean, unit-variance complex entries.}
 The matrix ${\bf \Theta}_{b,k}\in  \mathbb {C}^{N\times N}$ accounts for the UE specific channel correlation at BS $b$. {The pathloss due to large scale fading is implicitly considered in the correlation matrix unless otherwise stated. In the latter case, pathloss values are explicitly declared by expressing the correlation matrix as $a^2_{b,k}{\bf \Theta}_{b,k}$  where $a^2_{b,k}$ accounts for pathloss from BS $b$ to UE $k$. }The correlated scenario is 
motivated by the lack of space for implementing large antenna array and poor scattering environment~\cite{NgoMarzeta2011} that must be considered for a realistic performance evaluation. Moreover, the generic correlated model takes into account distinct angle of arrivals and angular spreads of UEs' signal for designing the precoder vectors. Also, it allows arbitrary configuration for the antenna array, including geographically distributed arrays.
\vspace{-0.5cm}
\section{{Large System Analysis}}
\label{sec: large system analysis}
In the following, we exploit the theory of large random matrices~\cite{RMT} to compute the so-called deterministic equivalents of the optimal Lagrangian multipliers $\{\lambda_k^*\}$ given by~\eqref{eq:lambda itr} under the generic channel model presented in Section~\ref{sec:Ch model}. Plugging such deterministic equivalents into~\eqref{eq:G_matrix} allows characterization of the coupling matrix elements $\{[{\bf G}]_{i,j}\}$ in \eqref{eq:G_matrix} in asymptotic regime, which consequently gives the asymptotically optimal scaling factors $\{\delta_k\}$. In doing so, the following assumptions (widely used in the literature) are made to properly define the growth rate of system dimensions,
\begin{assumption}\label{as:0}
As $N\to \infty$, $\!0 < \!\lim \inf  \frac{K}{N} \le \lim \sup  \frac{K}{N} < \infty$.
\end{assumption}
\begin{assumption}
	\label{as:1}
	The spectral norm of ${\bf \Theta}_{b,k}$ is uniformly bounded as $N\to\infty$, i.e., $
\!	\lim \sup_{{N \to \infty}}$ $ \!\!\!\max_{\forall b,k} \{\left\|{\bf \Theta}_{b,k}\right\|\}\!\! < \infty.$
\end{assumption}

\subsection{{Deterministic Equivalents For Lagrangian Multipliers}}
\label{sec:uplink asympto}
The derivation of the deterministic equivalents for the Lagrangian multipliers needs special handling because of their implicit formulation in~\eqref{eq:lambda itr}. In particular, dependency of $\boldsymbol{\lambda}^*$ on channel vectors prevents using trace lemma~\cite[Theorem 3.4]{RMT} explicitly for the denominator of~\eqref{eq:lambda itr}. The work in~\cite{LucaCouilletDebbahJournal2015} tackles this problem under i.i.d Rayleigh fading channels relying on a method introduced originally in~\cite{couillet2014large} in a different context. By using the same approach, the following result is obtained.
\begin{theorem}\label{th:up powers updw duality}
	Let  Assumptions \ref{as:0} and \ref{as:1} hold. If \eqref{Opt_problem} is feasible and its optimal solution is $\boldsymbol{\lambda}^*$, we have ${\rm{max}}_{\,\,k\,\,} |\lambda_k^*-\bar{ \lambda}_k| \rightarrow 0$ almost surely where
	\begin{align}
	\label{eq:lambdaebn}
	\bar \lambda_{k} = \frac{\gamma_{k}}{\bar{m}_{b_{k},k}}\quad \forall k\in \mathcal{U}
	\end{align}
and $\bar{m}_{b_k,k}$ is obtained as the unique non-negative solution of the following system of equations, evaluated for $b\in\mathcal{B}, i\in \mathcal{U}$
\vspace{-0.3cm}
	\begin{align}
	\label{eq:ST Th1}
	\bar{m}_{b,i} =\!\!{\rm{Tr}}\biggr({\bf \Theta}_{b,i}\biggr(\sum_{j\in \mathcal U}\frac{{\gamma_{j}}{\bf \Theta}_{b,j} }{{\bar{m}_{b_{j},j}}+\gamma_{j}{\bar{m}_{b,j}}} + \mu_b N {\bf I}_{N}\biggr)^{\scriptstyle-1}\biggr).
	\end{align}
\end{theorem}
\begin{IEEEproof} The proof is given in Appendix~\ref{sec:proof Theorem1}.\end{IEEEproof}
We observe that $\bar{ \lambda}_k \bar{m}_{b_k,k}$ represents the deterministic equivalent of the received SINR at BS $b_k$ when the MMSE receiver is aligned toward UE $k$. The UEs interact through the quantities $\bar{m}_{b,k}, \forall b,k$ such that, at the optimum, the SINR constraints for all UEs are asymptotically satisfied.
\vspace{-1cm}
\subsection{{Deterministic Equivalents For Coupling Matrix Entries}}
The deterministic equivalents for $\{\lambda_k^*\}$ in Theorem~\ref{th:up powers updw duality} are used to compute the asymptotically optimal receive beamforming vectors $\bar{\mathbf{v}}_{k}=(\sum_{j\in \mathcal U\setminus k}{\bar \lambda_{j}}{\mathbf{  h}}_{b_{k},j}{\mathbf{  h}}_{b_{k},j}\herm + \mu_{b_k}N\mathbf{I}_{N})^{-1}\mathbf{  h}_{b_k,k},\forall k$ in the dual uplink problem. By plugging $\{\bar{\bf{v}}_k\}$ into~\eqref{eq:G_matrix}, the following result is obtained.
\begin{theorem}\label{th:down pw updwn duality}
	Let Assumptions \ref{as:0} and \ref{as:1} hold, and assume \eqref{Opt_problem} to be feasible. Then, given the set of $\bar{ {\lambda}}_k$ and $\bar{m}_{b,k}, \forall k \in \mathcal{U},b\in \mathcal{B}$ as in Theorem~\ref{th:up powers updw duality}, we have $\left[{ \mathbf{ G}}\right]_{k,i} - \left[\bar{ \mathbf{ G}}\right]_{k,i} \to 0$ almost surely with
	\begin{align}
\label{eq:nonnormal G}
\left[\bar{ \mathbf{ G}}\right]_{k,i}= \begin{cases}
{ \gamma_{k}}/{\bar \lambda_k^2}& \textrm{for} \,\,\, i=k \\
-\frac{1}{N}\frac{\bar{m}_{b_i,i,k}^{\prime}}{\left(1 + \bar {\lambda}_k \bar m_{b_i,k}\right)^{2}}& \textrm{for} \,\,\, i\ne k
\end{cases}
\end{align}
where we have that $ [\bar{m}_{b,1,k}^{\prime},...,\bar{m}_{b,K,k}^{\prime}]=(\mathbf{I}_K-\mathbf{L}_{b})^{-1} {\vec u}_{b,k},\,\forall k\in\mathcal{U}$ and where	
\begin{equation}\label{eq:Proof en prime sys of equations3}
\left[\mathbf{L}_{b}\right]_{i,j}= \frac{1}{N^2}  \frac{  {\rm Tr}\left(\boldsymbol{\Theta}_{b,i} \mathbf{T}_{b}\boldsymbol{\Theta}_{b,j}\mathbf{T}_{b}\right) }{(1/\bar \lambda_j+  \bar m_{b,j})^2}
\end{equation}
and
\begin{equation}\label{eq:Proof en prime sys of equations4}
\begin{aligned}
{\vec u}_{b,k}=\!\!\left[\frac{1}{N} {\rm Tr} \left(\boldsymbol{\Theta}_{b,1} \mathbf{T}_{b}\boldsymbol{\Theta}_{b,k} \mathbf{T}_{b}\right),\ldots,\frac{1}{N} {\rm Tr}(\boldsymbol{\Theta}_{b,K} \mathbf{T}_{b}\boldsymbol{\Theta}_{b,k} \mathbf{T}_{b})\right]
\end{aligned}
\end{equation}
with ${\bf{T}}_{b}$ given by

\begin{align}\label{6.16}
{\bf{T}}_{b} = \left(\frac{1}{N}\sum\limits_{j\in \mathcal U}\frac{ \bar \lambda_j{\bf{\Theta}}_{b,j}}{1+ \bar \lambda_j\bar m_{b,j}} + \mu_{b}{\bf I}_N\right)^{-1}\!\!\!.
\end{align}	
\end{theorem}
\begin{IEEEproof} The proof is given in Appendix~\ref{sec:proof Theorem2}.\end{IEEEproof}
The term $\bar{m}_{b,i,k}'$ is the derivative of $\bar{m}_{b,i,k}(x)= \frac{1}{N}{\rm {Tr}}\big({\bf \Theta}_{b,i}(\frac{1}{N}\sum_{j\in \mathcal U}\frac{{\gamma_{j}}{\bf \Theta}_{b,j} }{{\bar{m}_{b_{j},j}}+\gamma_{j}{\bar{m}_{b,j}}} -x {\bf \Theta}_{b,k} + \mu_b {\bf I}_{N})^{-1}\big)$ with respect to the auxiliary variable $x$ and then evaluated at point $x=0$. The term $\bar{m}_{b_i,i,k}'$ in~\eqref{eq:nonnormal G} determines the coupling between UE~$i$ served by BS $b_i$ and UE $k$, and consequently indicates the level of interference leaking in between these two UEs.
The deterministic equivalents of entries $\{\left[\bar{ \mathbf{ G}}\right]_{k,i}\}$ can be used to compute the asymptotically optimal scaling factors as $\bar{\boldsymbol{\delta}} = \sigma^2 {\bar{\mathbf{G}}}^{-1} \mathbf{1}_{K}$, which depend only on the statistics of channel vectors. {Based on these results, we can now derive the deterministic equivalents of ICI terms $\{\epsilon_{b,k}\}$, and consequently present the coordination algorithms.}

\section{Distributed Optimization}
\label{sec:distributedOpt}
{The large system analysis provided in Section~\ref{sec: large system analysis}, gives the optimal power allocations in the asymptotic regime. Relying on these results, one can directly obtain 
asymptotically optimal receive and transmit beamforming vectors for UE $k$ as  
$\bar{\mathbf{v}}_{k}=(\sum_{j\in \mathcal U\setminus k}{\bar \lambda_{j}}{\mathbf{  h}}_{b_{k},j}{\mathbf{  h}}_{b_{k},j}\herm + \mu_{b_k}N\mathbf{I}_{N})^{-1}\mathbf{  h}_{b_k,k}$ and $\bar{{\vec w}}_{k}=\sqrt{\bar{\delta}_{k}/N} \bar{\vec v}_{k}$, respectively. The computation of asymptotic beamformers for UE $k$ needs only locally measured CSI at the serving BS $b_k$, i.e., ${\mathbf{  h}}_{b_{k},j},\,\forall j\in \mathcal{U}$, along with the asymptotic power allocations (i.e., $\{\bar \lambda_{j}\}$ and $\{\bar{\delta}_{k}\}$) whose computation needs only statistical information from neighboring BSs.
However, the resulting beamforming vectors satisfy the SINR constraints only asymptotically~\cite{LucaCouilletDebbahJournal2015} but not for a finite value of $N$, as demonstrated later by numerical examples. In order to ensure the SINR constraints, we invoke the solution briefly introduced in Section~\ref{sec:distibuted Sol intro}, which is exploited in more details in the following.  }
\vspace{-0.3cm}
\subsection{Distributed QoS Guaranteed Precoding}
{ The optimization problem formulation in~\eqref{Opt_problem} identifies the ICIs as the principal coupling parameters among BSs. Starting from the ICI formulation in~\eqref{3.1} and relying on the large system analysis in Section~\ref{sec: large system analysis}, we propose deterministic equivalents for these coupling parameters as
\begin{align}\label{3.4}
\bar \epsilon_{b,k}= -\bar \delta_{j} [\bar{\mathbf{G}}]_{k,j}
\end{align} 
with $\bar{\boldsymbol{\delta}} = \bar{ \mathbf{G}}^{-1} \mathbf{1}_{K}$, and the deterministic equivalents for the elements of coupling matrix $\{[\bar{\mathbf{G}}]_{k,j}\}$ given as in Theorem~\ref{th:down pw updwn duality}.
 Observe that the computation of \eqref{3.4} requires only the channel correlation matrices $\{{\bf \Theta}_{b,k},\,\forall b,k\}$ to be shared among BSs. Thus, plugging the deterministic ICIs from~\eqref{3.4} into the optimization problem in~\eqref{Opt_problem}, the centralized problem can be decoupled into independent sub-problems at each BS. This solution is summarized in Algorithm~\ref{alg:ICI_approx}. }
 \begin{algorithm} [H]
 	\caption{Decentralized beamforming with approximate ICI values.}
 	\label{alg:ICI_approx}
 	
 	\begin{algorithmic}[1]
 		\LOOP
 		\IF{Any change in the UEs' statistics or during the initial stage}
 		\STATE Each BS sends the updated correlation matrices to the coupled BSs via backhaul.
 		\STATE Update $\bar{\boldsymbol{\lambda}}$ and $\bar{\boldsymbol{\delta}}$ values locally based on Theorem~\ref{th:up powers updw duality} and~\ref{th:down pw updwn duality}.
 		\STATE Update the approximate ICIs locally based on~\eqref{3.4}.		\ENDIF
 		\STATE Use the approximate ICIs as fixed $\epsilon_{b,k}$ in \eqref{Opt_problem}, and solve the sub-problems locally to get the downlink beamformers.
 		\ENDLOOP
 	\end{algorithmic}
 \end{algorithm}
\vspace{-0.2cm}
 {  Algorithm~\ref{alg:ICI_approx} allows the BSs to obtain the precoders locally relying only on shared statistics and locally available CSI. The resulting precoders satisfy the SINR constraints.  The sub-problems at each BS can be solved using a convex optimization solver or fixed point iterations as shown in~\cite{Pennanen-Tolli-Latva-aho-TSP-13}.  Note that the resulting problem with approximate ICIs is a restriction of the original problems~\eqref{eq:prim problemSimple} and~\eqref{Opt_problem}, and the infeasibility rate and total transmission power would increase, depending on the accuracy of the ICI approximations. However, since the
	deterministic equivalents of ICI terms provide good approximations for the optimal
	ICI values in the finite regime, the performance loss is small for a relatively moderate number of UEs and antennas.
}
\vspace{-0.3cm}
\subsection{{	An Alternative Distributed Precoding Method With Reduced Backhaul Signaling}}
   Although each ${\bf \Theta}_{b,k}$ changes slowly in time compared to small-scale fading components, the exchange of such information among coupled BSs via backhaul links may not be practical when $N$ and $K$ are large. To overcome this issue, a heuristic solution is proposed that reduces the amount of shared information subject to slightly higher transmit power (as shown in numerical results). We notice that  $[\bar{\mathbf{G}}]_{k,j},\forall j\in \mathcal U_{b}$ are the coupling terms between UE $j,\forall j\in \mathcal U_{b}$ and UE $k$, and define the amount of interference leaking from the precoders of UE $j,\forall j\in \mathcal U_{b}$ to UE $k$. In particular, observe that the amount of interference from an interfering BS $b$ to a given UE $k$ in~\eqref{3.4} is given in terms of $[\bar{\mathbf{G}}]_{k,j},\forall j\in \mathcal U_{b}$ and $\delta_{j},\,\forall j\in \mathcal U_{b}$. On the other hand, for a given set of Lagrangian multipliers $\{\bar{{\lambda}}_k\}$, the coupling terms $[\bar{\mathbf{G}}]_{k,j},\,\forall j\in \mathcal U_{b}$ in~\eqref{eq:nonnormal G} depend only on the statistics locally available at the interfering BS~$b$.
    This observation motivates extraction of an approximation for the interference that a BS $b$ causes to a non-served UE $k$ based on partial knowledge of non-local statistics (statistics available at other BSs) while utilizing the locally available statistics. 
In doing so, we declare the large-scale attenuation (due to pathloss and fading) explicitly and express the correlation matrices as $a^2_{b,k}{\bf \Theta}_{b,k}$  where $a^2_{b,k}$ accounts for pathloss from BS $b$ to UE $k$. We assume that each BS $b$ is able to estimate (perfectly) the channel correlation matrices $a^2_{b,k}{\bf \Theta}_{b,k},\, \forall k$ 
while the correlation matrices $a^2_{b',k}{\bf \Theta}_{b',k}, \, \forall k$ from all other BSs $b' \neq b$ are not known locally at BS~$b$.
Only the large-scale attenuation values $\{a^2_{b',k}\}$ are assumed to be available for the non-local channels.
The first assumption relies on the observation that correlation matrices remain constant for a sufficiently large number of reception phases to be accurately estimated at the BS~\cite{BjörnsonLuca2016}. The second one is motivated by the observation that most current standards require the UEs to periodically report received signal strength indication (RSSI) values to their serving BSs (usually using orthogonal uplink resources). Under the assumption that nearby BSs are also able to receive such RSSI measurements, a partial knowledge of non-local channel statistics can be obtained without any information exchange through backhaul~links\footnote{Alternatively, RSSI values can be exchanged among BSs over backhaul links.}. 
    \begin{figure*}[tpb]
	\begin{center}
		\includegraphics[clip, trim=0cm 0cm 0cm 0cm, width=\textwidth]{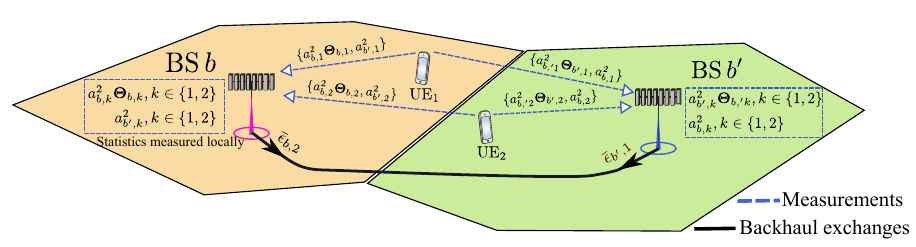}
	\end{center}\vspace{-0.3cm}
	\caption{ {An illustration of locally measured statistics and backhaul signaling in Algorithm 2.}}
	\label{fig:Heuristic algor}
\end{figure*}

Under the above assumptions, each BS can locally compute (through Theorems \ref{th:up powers updw duality} and \ref{th:down pw updwn duality}) approximations of the optimal powers along with the coupling parameters, which can be used in \eqref{3.4} to locally obtain an approximation for the interference that the BS causes to a non-served UE. The approximate ICI values are then sent to the respective serving BSs over the backhaul link to be plugged in the ICI constraints of the local optimization problems. {Fig.~\ref{fig:Heuristic algor} shows a two-cells example where a given BS $b$ locally measures the statistics  of the local channel vectors, i.e., ${a}_{b,k}^2 {\bf \Theta}_{b,k}, k\in\{1,2\}$, and obtains the pathlosses of the non-local channels i.e., ${a}_{b',k}^2, k\in\{1,2\}$, from the reported received signal strength indication (RSSI). Finally, BS $b$ sends the approximation for the interference that it causes to UE2 over the backhaul link. 
This solution is summarized in Algorithm \ref{alg:ICI_approx heuristic}.}
\vspace{-0.2cm}
\begin{algorithm} [h]
		\caption{Heuristic solution}
	\label{alg:ICI_approx heuristic}
	\begin{algorithmic}[1]
		
		\LOOP
		\IF{Any change in the UEs' statistics or during the intial stage}
		\STATE Users broadcast the pathloss information to the nearby BSs using uplink resources.
		\STATE Each BS $b$ locally calculates approximations for $\delta_k$ and $[{\bf G}]_{i,j}$ values using Theorems~\ref{th:up powers updw duality} and~\ref{th:down pw updwn duality}  where  BS $b$ locally assumes ${a}_{b',k}^2 {\bf \Theta}_{b',k}={a}_{b',k}^2{\bf I}_{N} \ \forall k$ for all $b'\neq b$.
		\STATE  ICI values $\bar{\epsilon}_{b,k}, \forall k\not\in \mathcal{U}_{b}$ are computed from \eqref{3.4} at each BS $b$.
		\STATE  Each BS $b$ sends the ICI values $\bar{\epsilon}_{b,k}, \forall k\not\in \mathcal{U}_{b}$  to the corresponding serving BSs.
		\ENDIF
		\STATE  BSs use the approximate ICIs as fixed $\epsilon_{b,k}$ in~\eqref{Opt_problem} and solve the sub-problems locally to get the downlink precoders.
		\ENDLOOP
	\end{algorithmic}
\end{algorithm}
\vspace{-0.5cm}
\subsection{Backhaul Signaling And Complexity Analysis}
  {    Table~\ref{tab:t1} presents the locally available CSI and the required information exchange over the backhaul links for a given BS~$b$ to obtain the beamformers with Algorithms 1 and 2.
	Algorithm~\ref{alg:ICI_approx heuristic} is a semi-static coordination method that, unlike the available decoupling methods requiring a continuous exchange of CSI messages~\cite{ Tolli-Pennanen-Komulainen-TWC10, Pennanen-Tolli-Latva-aho-spl-11, ChaoADMM2012, Pennanen-Tolli-Latva-aho-TSP-13}, relies only on local channel statistics and reported path gain values. This makes it more resilient to limited link capacity and latency.
	Unlike Algorithm~\ref{alg:ICI_approx} that needs exchanges of correlation matrices over the backhaul links, Algorithm~\ref{alg:ICI_approx heuristic} sends the approximate ICI values (scalars) on the backhaul links only when sufficient changes occur in the channel statistics. This reduces the exchange rate by almost $N^2/2$.
	In the numerical analysis of Section~\ref{sec:Simulation Results}, a small difference in the transmission powers of these algorithms is observed, which is due to the difference in the accuracy of approximate ICI values.}

{Concerning the complexity analysis, we notice that the proposed algorithms need to solve the sub-problems at BSs subject to the fixed ICIs given by~\eqref{3.4}. The solution to such sub-problems can be obtained using SOCP, semidefinite programming (SDP) and uplink-downlink duality~\cite{Pennanen-Tolli-Latva-aho-TSP-13}. This latter approach requires lower computational complexity compared to SOCP and SDP~\cite{Pennanen-Tolli-Latva-aho-TSP-13}. Specifically, at a given BS $b$, the Lagrangian multipliers associated with rate and ICI constraints are evaluated via a projected sub-gradient method and a simple fixed point iteration~\cite{Pennanen-Tolli-Latva-aho-TSP-13}. This involves a matrix inversion of size $N\times N$ with a complexity per iteration in the order of $\mathcal{O}(|\mathcal{U}_b|\times N^3)$ where $|\mathcal{U}_b|$ denotes the number of  UEs served by BS $b$.
	Concerning the calculation of approximate ICIs, we notice that the ICI terms in~\eqref{3.4}
	are updated only when there are sufficient changes in channel matrix statistics, which vary at a
	much slower rate than the fading CSI. 
     The computation of approximate ICIs requires evaluation of $\{{\bar{\lambda}}_k\}$ and $\{{\bar{\delta}}_k\}$ values with a complexity of order $\mathcal{O}(K\times N^3)$ and $\mathcal{O}(K^3)$ respectively.}
\begin{table}[]
	\centering
	\caption{Locally available knowledge and acquired information over backhaul at BS $b$}
	\label{tab:t1}
	{
		\begin{tabular}{l|l|l|}
			\cline{2-3}
			& \scriptsize{Local CSI at BS $b$} &  \scriptsize{Acquired information from other BSs} \\ \hline
			\multicolumn{1}{|l|}{\!\!\!\scriptsize{Alg. 1}\!\!\!} & \!\!$\{{a}_{b,j}^2 {\bf \Theta}_{b,j}\}, \{{\bf{h}}_{b,j}\}, \forall j\in\mathcal{U}$\!\! & $\{{a}_{b',j}^2 {\bf \Theta}_{b',j}\}, \forall j\in\mathcal{U},\forall b'\neq b$\\ \hline
			\multicolumn{1}{|l|}{\!\!\!\scriptsize{Alg. 2}\!\!\!} & \!\!$\{{a}_{b,j} ^2{\bf \Theta}_{b,j}\}, \{{\bf{h}}_{b,j}\}, \forall j\in\mathcal{U}$\!\! & $\{\bar{\epsilon}_{b',j}\}, \forall j \in\mathcal{U}_{b},\forall b'\neq b$ \\ \hline
	\end{tabular}}
\vspace{-0.5cm}
\end{table}

\section{Network with Partitioned UE Population}
{ So far, we have assumed distinct statistical properties for UEs. However, as in many works in literature (such as~\cite{UserPartionAdhikary,GroupKim2015}), one can consider a special scenario where the UEs are grouped on the basis of their statistical properties. In particular, each BS partitions the UE population into groups such that the eigenspaces of correlation matrices in distinct groups be asymptotically orthogonal  (this is referred to as asymptotic orthogonality condition).
The main idea of this section consists of exploiting the asymptotic orthogonality condition, and the similarities of statistical properties of nearly co-located UEs~\cite{UserPartionAdhikary}  to get both mathematically and computationally simpler approximations of the ICI terms. The dependency of approximate ICIs on group-specific correlation properties allows further reduction in the backhaul exchange rate of the decentralized solutions in Section~\ref{sec:distributedOpt}. Moreover, the analysis motivates development of the decentralization framework within a context similar to two-stage beamforming~\cite{UserPartionAdhikary} (discussed further in Section~\ref{sec:Conclusion}).

{The UE-grouping idea is presented next by using the simple two-cell configuration in Fig.~\ref{fig:Groups}. The detailed multicell model is presented mathematically later on. Consider a BS, equipped with a linear array, which properly partitions the UE population into distinct groups such that the angle of arrivals of UEs' signals within distinct groups, at the given BS, are sufficiently separated. Assuming one-ring channel model\footnote{In a typical cellular configuration with a tower-mounted BS and no significant local scattering, the propagation
        	between the BS antennas and any given UE is expected to be characterized by the local scattering around the UE resulting in the well-known one-ring model~\cite{shiu2000fading}. }, it is shown in~\cite{UserPartionAdhikary} that the correlation matrices of UEs in distinct groups have nearly orthogonal eigenspaces as $N\rightarrow \infty$.        
Following this line of thoughts, BS $b$ ($b'$) in Fig.~\ref{fig:Groups}, equipped with a linear array, resolves two smaller groups $g_1,g_2$ (equivalently $g_4,g_5$ for BS $b'$) and one bigger group $g_3$ (equivalently $g_6$ for BS $b'$ ) with non-overlapping supports of AoA distributions. The beams in the figure represent the AoA spread of UEs' signal at BSs. Equivalently, the groups with dashed (dotted) contours correspond to the partition related to BS $b$ ($b'$). 
				In the following, the index $g$ is used to refer to the $g^{\text{th}}$ group, the set of UEs in group $g$ is denoted by $\mathcal{G}_g$ and the set of all groups at a given BS~$b$~is~denoted~by~$\mathcal{A}_b$.}

\label{sec:grouped UEs}
\begin{figure*}[tpb]
	\begin{center}
		\includegraphics[clip, trim=0cm 0cm 0cm 0cm, width=\textwidth]{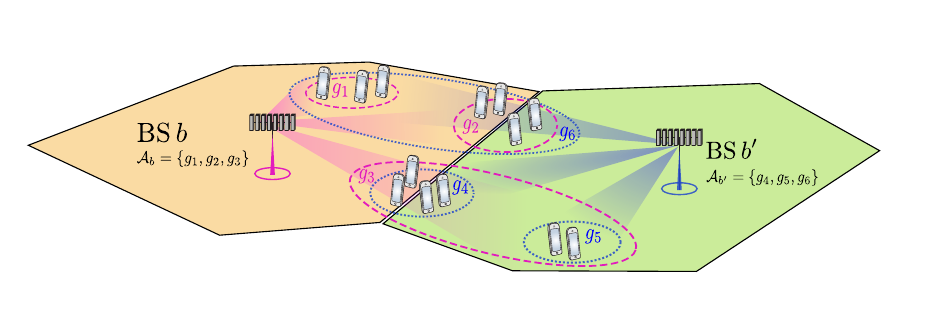}
	\end{center}\vspace{-0.3cm}
	\caption{{An illustration of UE-groups formation at BSs based on UEs' angular separation.
	}}
	\label{fig:Groups}
\end{figure*}
 {In the following, the aforementioned assumptions are presented mathematically for a generic multicell setting.
 		We declare the large-scale attenuation $a^2_{b,k}$ explicitly and express the correlation matrices as $a^2_{b,k}{\bf \Theta}_{b,k}$. Also, the correlation matrices are assumed to have eigenvalue decomposition given as $a^2_{b,k}\boldsymbol{\Theta}_{b,k}=a^2_{b,k}{\bf U}_{b,k}{\bf \Xi}_{b,k}{\bf U}_{b,k} \herm$.
 		The diagonal $r_{b,k}\times r_{b,k}$ matrix ${\bf \Xi}_{b,k}$ holds the non-zero eigenvalues, and the corresponding eigenvectors are stacked in ${\bf U}_{b,k}\in \mathcal{C}^{N\times r_{b,k}}$. Under grouping assumption, we declare the aforementioned asymptotic orthogonality condition as follows: ${\bf U}_{b,i}\herm {\bf U}_{b,j}= {\bf 0},\, \forall i\in \mathcal{G}_g, j\not\in \mathcal{G}_g,\forall g\in \mathcal{A}_b$ as $ N \rightarrow \infty$~\cite{UserPartionAdhikary}, which indicates the orthogonality of eigenspaces of correlation matrices for UEs in distinct groups in asymptotic regime. Within this context a mathematically attractive problem is to assume a 
  homogeneous model wherein the UEs in a group have identical correlation while experiencing different pathlosses i.e. $a_{b,k}^2 \boldsymbol{\Theta}_{b,k}=a_{b,k}^2\boldsymbol{\Theta}_{b,g}, \,\forall k \in \mathcal{G}_g,\forall g\in \mathcal{A}_b$ where $a_{b,k}^2$ accounts for UE specific pathloss values and $\boldsymbol{\Theta}_{b,g}$ dictates the group specific correlation properties. These conditions 
   are not verified exactly in practice. However, as UEs in the same group are likely to be nearly co-located, one might closely approximate these conditions by proper UE scheduling (specifically when there is a large pool of UEs to be scheduled), which is not however in the scope of this work. 

  Given the aforementioned system model, one can write  
 $\bar{m}_{b,i}$ in~\eqref{eq:ST Th1} for UEs in a group $g$ resolved at BS $b$ as $\bar{m}_{b,i}={a}_{b,i}^2\bar{\eta}_{b,g}$~with $\bar{\eta}_{b,g}$ being a group specific measure evaluated as $\forall g \in \mathcal{A}_b, \forall b$
 \begin{equation}
 \label{eq:eta multi cell}
 \bar{\eta}_{b,g} = \sum_{i=1}^{r_{g}} 
 {\biggr( \sum_{j\in \mathcal{G}_g} \frac{{\gamma_{j}}
 	}{      {a_{b_j,j}^2}\bar{\eta}_{b_j,g_{j}}/{a_{b,j}^2}+\gamma_{j}\bar{\eta}_{b,g}}  +  {N\mu_b}({[{\bf \Xi}_{b,g}]_{i,i}}) \biggr)^{-1}}
 \end{equation}
 where 
for a given UE $j$, we use $g_j$ to denote the group resolved at the serving BS $b_j$, which contains UE $j$ as a member. 
The formulation of measure $\bar{\eta}_{b,g}$ follows directly from~\eqref{eq:ST Th1} by substituting $\boldsymbol{\Theta}_{b,k},\,\forall k \in \mathcal{G}_g$ with $a_{b,k}^2\boldsymbol{\Theta}_{b,g}, \,\forall k \in \mathcal{G}_g$. Note that $\boldsymbol{\Theta}_{b,g}={\bf U}_{b,g}{\bf \Xi}_{b,g}{\bf U}_{b,g} \herm$ with ${\bf U}_{b,g}\herm {\bf U}_{b,g'}= {\bf 0},\, \forall g,g'\in \mathcal{A}_b,\, g\neq g'\, {\text{ as}}\, N\rightarrow \infty$, and ${\bf U}_{b,g}\herm {\bf U}_{b,g}= {\bf I}_{r_{g}}$, which gives $\bar{\eta}_{b,g}$ as in~\eqref{eq:eta multi cell}.
The interactions among UE groups are regulated using $\bar{\eta}_{b,g}$ values in~\eqref{eq:eta multi cell} such that the SINR constraints for all UEs within the groups are satisfied asymptotically. 
The corresponding asymptotically optimal uplink power for UE $j$ can be evaluated as $\bar{\lambda}_j/N=\gamma_j/(N a_{b_j,j}^2\bar{\eta}_{b_j,g_j})$. 

As a consequence of the considered system model, the optimal ICI terms, as the inter-cell coordination messages, can be characterized directly in terms of channel statistics and total transmit power per group as stated in the following corollary.
 \vspace{-0.2cm}
 \begin{corollary}
 	\label{cor:homo closeICI}
 	Consider the multicell scenario with grouped UEs experiencing homogeneous correlation properties $a_{b,k}^2\boldsymbol{\Theta}_{b,k}=a_{b,k}^2\boldsymbol{\Theta}_{b,g}, \,\forall k \in \mathcal{G}_g,\forall g\in \mathcal{A}_b$. {Then under Assumption \ref{as:1} and given the growth rate defined in Assumption~\ref{as:0}, the optimal ICI values $\epsilon_{b,k},\forall k \in \mathcal{G}_g\centernot\cap \mathcal{U}_b$ converge almost surely to the deterministic equivalents $\bar{\epsilon}_{b,k}$ with}
 		\vspace{-0.2cm}
 	\begin{equation}
 	\label{eq:determ ICI homogen}
 	\bar \epsilon_{b,k}=  \varphi_{b,k} {a_{b,k}^2} \bar{P}_{b,g},\,\forall k \in \mathcal{G}_g\centernot\cap \mathcal{U}_b
 	\end{equation}
 	where
 	\vspace{-0.3cm}
 	\begin{equation}
 	\label{eq:varphi in ici}
 	\varphi_{b,k}= \frac{{\rm Tr}  ((\boldsymbol{\Theta}_{b,g} \mathbf{T}_{b,g})^2)/{\rm Tr} ( \boldsymbol{\Theta}_{b,g} \mathbf{T}_{b,g}^2)}{(1+\gamma_k(a_{b,k}^2 \bar \eta_{b,g})/(a_{b_k,k}^2 \bar \eta_{b_k,g_k}))^2}
 	\end{equation}
 	where the term $\bar \eta_{b,g}$ is the group specific parameter given by~\eqref{eq:eta multi cell} or equivalently as $\bar \eta_{b,g}=\frac{1}{N}{\rm Tr}({\boldsymbol \Theta}_{b,g}{\mathbf T}_{b,g})$ with ${\bf{T}}_{b,g} = (\frac{1}{N}\sum_{j\in \mathcal{G}_g}({ \gamma_{j}{a_{b,j}^2\bf{\Theta}}_{b,g}})/({a_{b_j,j}^2 \bar \eta_{b_j,g_j}+ \gamma_j a_{b,j}^2 \bar \eta_{b,g}}) + \mu_{b}{\bf I}_N)^{-1}$. The asymptotic total transmit power at BS~$b$ required for serving UEs in $\mathcal{G}_g\cap \mathcal{U}_b$ is denoted by $\bar{P}_{b,g}$.
 	Stacking all $\bar{P}_{b,g},\forall g\in\mathcal{A}_b, \forall b$ in a vector, we get
 	the system of equations $[\bar{P}_{1,1},...,\bar{P}_{1,|\mathcal{A}_1|},...,\bar{P}_{|\mathcal{B}|,|\mathcal{A}_{|\mathcal{B}|}|}]=({\bf D}-{\bf L})^{-1}\bf u$ where the elements of vector $\bf u$ and matrix $\bf L$ are given as $[{\bf u}]_{(b,g)}=\frac{1}{N}\sum_{j\in \mathcal U_{{{b}}}\cap \mathcal{G}_g} {\gamma_j}/{ a_{{{b}},j}^2 }$  and $[{\bf L}]_{(b',g'),(b,g)}=\frac{1}{N}\sum_{j\in \mathcal U_{{b'}}\cap \mathcal{G}_g\cap \mathcal{G}_{g'}}  \varphi_{b,j} {\gamma_j{a_{b,j}^2} }/{{a_{{b'},j}^2} }$, respectively. The tuple index $(b',g')$ points to the row/column element corresponding to group $g'$ at BS $b'$. The matrix
 	$\bf D$ is a diagonal matrix where $[{\bf D}]_{(b,g),(b,g)}=\frac{\bar{\eta}_{{{b,g}}}^2} {\bar{\zeta}_{{{b,g}}}^{\prime}}$ with 
 	 $ 
 	\bar{\zeta}_{b,g}^{\prime}=\frac{1}{N}{\rm Tr  \big(\boldsymbol{\Theta}_{b,g} \mathbf{T}_{b,g}^2}/{(1-\rho_{b,g} \rm Tr  ((\boldsymbol{\Theta}_{b,g} \mathbf{T}_{b,g})^2)\,)\big)}
$
 	and $\rho_{b,g}=\frac{1}{N^2} \sum_{j \in \mathcal{G}_g} 1/{(
 		\frac{a_{b_j,j}^2}{\gamma_j a_{b,j}^2}\bar{\eta}_{b_j,g_j}
 		+ \bar{\eta}_{b,g})^2}$.
 \end{corollary}
 \begin{IEEEproof}
 	The proof is given in Appendix~\ref{subse:proof of corr ICI} .
 \end{IEEEproof}
 Looking at the derived expression in~\eqref{eq:determ ICI homogen}, we observe that the asymptotically optimal ICI $\bar{\epsilon}_{b,k}$ is directly related to the group aggregated transmit power at the interfering BS $b$ degraded by the pathloss $a_{b,k}^2$. 
The total transmit power is generally proportional to the ratio $K/N$~\cite{ICC17Asghari}. Therefore, ICIs are expected to go to zero only when there exists a large imbalance between these quantities. 
The parameter $\varphi_{b,k}$ in~\eqref{eq:varphi in ici} indicates that the UEs with higher SINR targets are generally assigned smaller ICIs. Also, the target SINRs are multiplied by $(a_{b,k}^2 \bar \eta_{b,g})/(a_{b_k,k}^2 \bar \eta_{b_k,g_k})$ terms to reflect the position of the UEs with respect to the serving and the interfering BSs, as well as the priority of the BSs via $\bar \eta_{b,g}$ terms.

 Assuming properly partitioned UE population, one can directly utilize the ICI expression from Corollary~\ref{cor:homo closeICI} to obtain the approximate ICIs in Algorithm~\ref{alg:ICI_approx}. This brings two benefits: Firstly, the approximate ICIs can be attained based on group-specific correlation properties that reduces the backhaul exchange rate. Secondly, the computational effort for ICI evaluation is decreased.  Denoting the total number of groups formed at all BSs by $M$, we notice that the computation of asymptotic ICIs for UEs of a group in~\eqref{eq:determ ICI homogen} requires an $M\times M$ matrix inversion as compared to $K\times K$ matrix inversion required in the generic formulation~\eqref{3.4}. Generally, we expect $M$ to be much smaller than  $K$. The viability of this approach is studied numerically in Section~\ref{sec:Simulation Results}.

\section{Numerical Analysis}
\label{sec:Simulation Results}
Monte Carlo simulations are now used to validate the performance of the proposed solutions. The performance metrics are averaged over 1000 independent UE drops and channel realizations. {We consider a network with $L$ cells and assume that the same number of $\bar{K}=\frac{K}{L}$ UEs is assigned to each cell. The UEs are distributed uniformly in the coverage area of cells.}
The pathloss function is modeled as~$a_{b,k}^2=({d_{0}}/{d_{b,k}})^{3}$ where $d_{b,k}$ represents the distance between BS $b$ and UE $k$, and $d_{0}=1$ m is the reference distance. The BSs are placed 1000m apart from each other. The transmission bandwidth is $W = 10$ MHz, and the total noise power $\sigma^2=WN_0$ is -$104$ dBm. By assuming a diffuse 2-D field of isotropic scatterers around the receiver~\cite{JakesMicrowavebook}, the correlation matrix for an antenna element spacing of $\Delta$ is given by
\begin{equation}\label{eq:corr model}
\left[{\bf{\Theta}}_{b,k}\right]_{j,i}=\frac{a_{b,k}^2}{\varphi_{b,k}^{\max}-\varphi_{b,k}^{\min}}\int_{\varphi_{b,k}^{\min}}^{\varphi_{b,k}^{\max}} \! e^{i\frac{2\pi}{w}\Delta (j-i){\rm cos}(\varphi)} \, \mathrm{d}\varphi
\end{equation}
 where waves arrive with an angular spread $\Delta \varphi$ from $\varphi_{\min}$ to $\varphi_{\max}$. The wavelength is denoted by $w$, and the antenna element spacing is fixed to half the wavelength $\Delta=1/2w$.

 \subsection{{Performance Evaluation Of The Distributed Precoding Methods}}
Next, the analyses of Sections~\ref{sec: large system analysis} and~\ref{sec:distributedOpt} are validated numerically. The UEs are dropped in a network with $7$ cells.
Fixed wide angular spread ($\Delta \varphi_{b,k} =\frac{\pi}{2}$) is assigned to UEs served by a given BS that accounts for well-conditioned correlation matrices. While the channels to non-served UEs (that are far from the given BS) have an angular spread of ${\pi}/{6}$ which yields rank deficient correlation matrices. The angle of arrival is determined by the angular position of the UEs with respect to the BSs.

%
%
{Theorems~\ref{th:up powers updw duality} and~\ref{th:down pw updwn duality} present the asymptotically optimal power assignments in downlink and dual uplink problems in terms of channel statistics, which is instrumental to get further insight into the structure of the optimal solution. As mentioned in Section~\ref{sec:distributedOpt}, one might
 serve UEs in finite system regime by utilizing
 the asymptotic power terms in Theorems~\ref{th:up powers updw duality} and~\ref{th:down pw updwn duality}, i.e., $\{\bar{\lambda}_{K}\}$ and $\{\bar{\delta}_{K}\}$, and corresponding receive/transmit beamforming vectors, i.e., $\bar{\mathbf{v}}_{k}=(\sum_{j\in \mathcal U\setminus k}{\bar \lambda_{j}}\mathbf{  h}_{b_{k},j}\mathbf{  h}_{b_{k},j}\herm + \mu_{b_k}N\mathbf{I}_{N})^{-1}\mathbf{  h}_{b_k,k}$ and $\bar{{\vec w}}_{k}=\sqrt{\bar{\delta}_{k}/N} \bar{\vec v}_{k}$, respectively. However, this approach only guarantees the target rates to be satisfied asymptotically. This is shown in Fig.~\ref{fig:ratevspowerUp} where the empirical CDF of the achievable rates is depicted in downlink and dual uplink problems.  The number of antennas is the same as the number of UEs, and the UEs are served with a target rate of 1 bit/s/Hz/UE using the asymptotically optimal beamformers, i.e., $\bar{{\vec w}}_{k}$ and $\bar{\vec v}_{k}$.
It can be seen that the achievable rates are mainly concentrated around the target rates. However, as shown in Fig.~\ref{fig:ratevspowerUp}-b, 30 percent of UEs attains a rate of less than 0.7 bit/s/Hz/UE in the downlink with $N = 14$ antennas. This percentage reduces to 12 percent when $N$ increases up to almost 100. The empirical CDF shows that the deviation from the target rate decreases as dimensions of the problem increase and the rate constraints for all UEs are expected to be satisfied asymptotically.}

\begin{figure*}
	\centering
	\begin{subfigure}{.51\textwidth}
		\centering
		\includegraphics[width=\columnwidth]{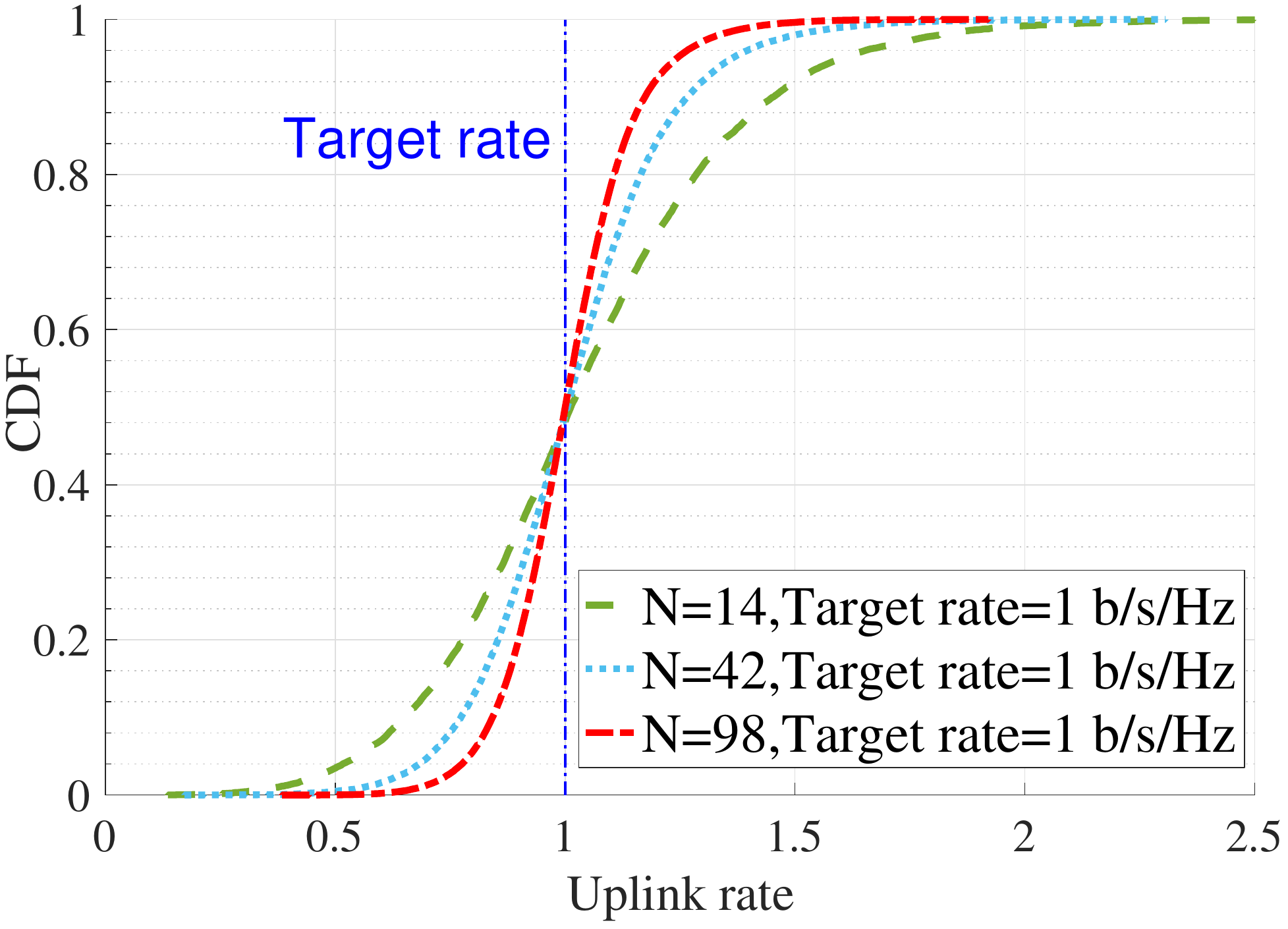}
				\caption{CDF of achievable uplink  rates.}
	\end{subfigure}%
	\begin{subfigure}{.49\textwidth}
		\centering
	\includegraphics[width=\columnwidth]{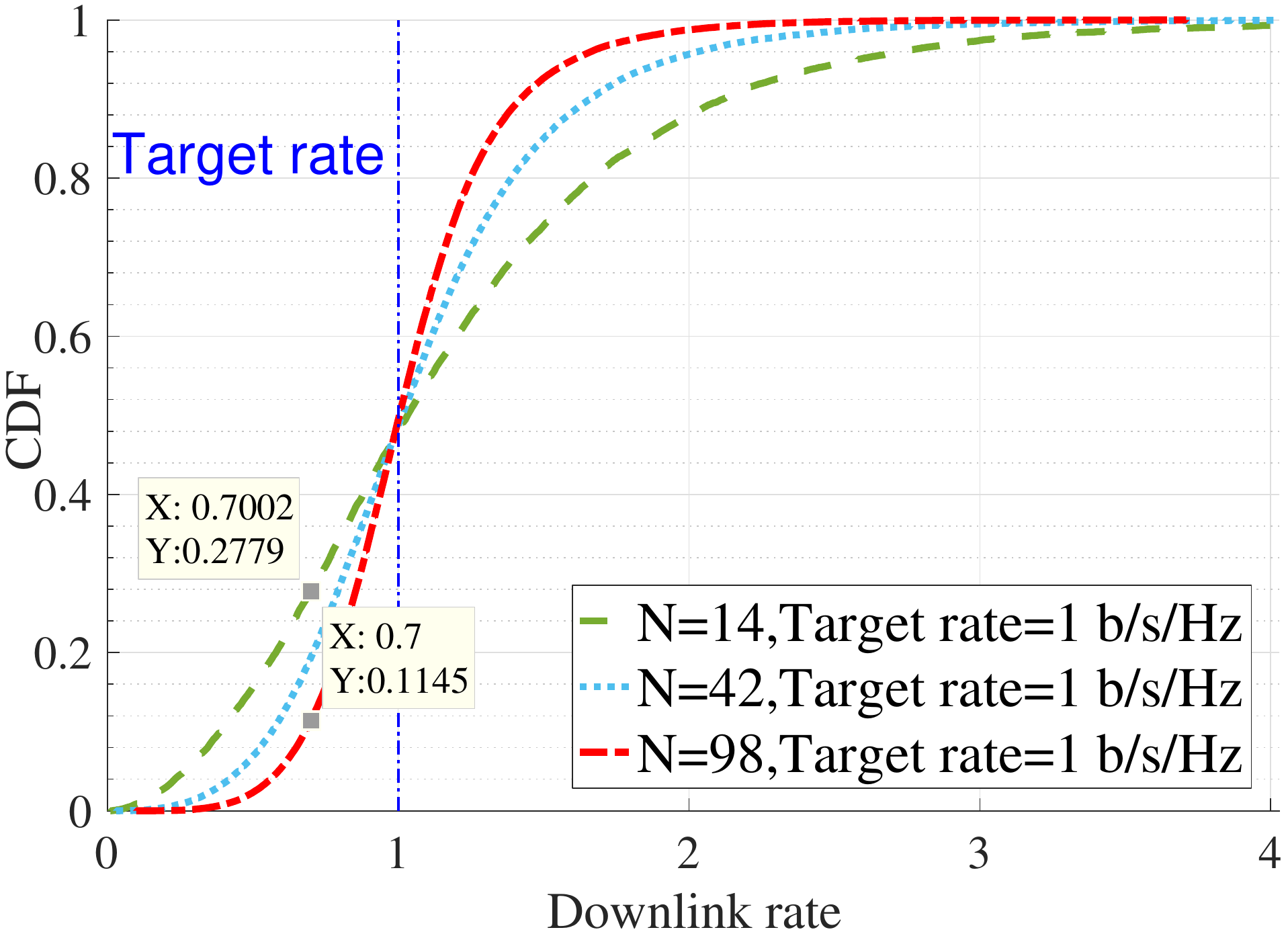}
		\caption{CDF of achievable downlink rates.}
	\end{subfigure}
	\caption{Empirical CDF of achievable rates using asymptotic beamformers, $\frac{N}{K}=1$.}
\label{fig:ratevspowerUp}
\end{figure*}
\vspace{0.3cm}

As mentioned in Section~\ref{sec:distributedOpt}, we can explore the availability of local CSI while relying on deterministic equivalents of ICI values to obtain the QoS guaranteed precoders. Both Algorithm~\ref{alg:ICI_approx} and~\ref{alg:ICI_approx heuristic} satisfy the SINR constraints while having a minimum cooperation among BSs.
Fig.~\ref{fig:perfcomp SNR1} presents the averaged total transmission power required for serving UEs with target rates fixed to 1 bits/s/Hz/UE. 
The total number of UEs in the network
grows at the same rate as the number of antennas such that $\frac{N}{K}=1$ in Fig.~\ref{fig:perfcomp SNR1}-a and $\frac{N}{K}=2$ in Fig.~\ref{fig:perfcomp SNR1}-b.
Thus, the spatial loading is fixed as the number of antennas is increased. The optimal total transmission power using~\eqref{eq:prim problemSimple} is the reference curve denoted as the centralized approach. 
It can be seen in both Fig.~\ref{fig:perfcomp SNR1}-a and~\ref{fig:perfcomp SNR1}-b that Algorithm~\ref{alg:ICI_approx} satisfies the target rates subject to small performance degradation even for relatively small $N$ and $K$. This gap diminishes further as the number of antennas and UEs are increased.

{Based on asymptotic ICI expressions, we observed that the interference caused by a BS to a non-served UE depends mainly on the local statistics. Utilizing this in Algorithm~2, the approximate ICIs are evaluated at the interfering BSs relying on local and partial non-local knowledge of channel statistics.	
The viability of this approach can be seen from the small
difference in the transmission powers of Algorithms~\ref{alg:ICI_approx} and~\ref{alg:ICI_approx heuristic} in Fig.~\ref{fig:perfcomp SNR1}.
 } A heuristic case (included for comparison) labeled as 'i.i.d fully decentralized' is also depicted in Fig.~\ref{fig:perfcomp SNR1}, where the correlation properties are ignored and the approximated ICI values are derived relying only on pathloss information. Ignoring the correlation properties when designing the precoders generally results in a large performance degradation as depicted in Fig.~\ref{fig:perfcomp SNR1}. 
{Another observation is the smaller performance gap among all methods in Fig.~\ref{fig:perfcomp SNR1}-b with $\frac{N}{K}=2$, compared to Fig.~\ref{fig:perfcomp SNR1}-a with $\frac{N}{K}=1$, which is due to the increase in the number of degrees of freedom (d.o.f) per UE. In particular, for any given number of antennas, the number of UEs in Fig.~\ref{fig:perfcomp SNR1}-a is twice as large as that in Fig.~\ref{fig:perfcomp SNR1}-b. In general, the performance gap among various methods diminishes when the ratio of d.o.f per UE increases.
	As the ratio of d.o.f per UE goes to infinity the differences disappear. In particular, this  is illustrated in~\cite{Asgharimoghaddam-Tolli-Rajatheva-ICC2014} (the conference counterpart of the current work) under i.i.d Rayleigh fading channel where the transmission powers of all methods converge as $N$ grows large, given a fixed $K$.}

The other sub-optimal solutions, including ICZF and ZF, design the precoders locally without cross cell coordination. In particular, ICZF sets ICIs equal to zero while handling the local interference optimally. ZF attains the precoders by forcing all (intra-cell and inter-cell) interference terms to zero. These methods are infeasible for some UE drops in the case with $\frac{N}{K}=1$, thus a fair comparison is not possible and the corresponding curves are omitted in Fig.~\ref{fig:perfcomp SNR1}-a . This is in fact due to 
narrow angular spreads, which result in lack of degrees of freedom for nulling the interference to UEs with overlapping angular spreads. In the case with $\frac{N}{K}=2$, ICZF and ZF attain the target rates subject to 4 dB higher transmission power as compared to Algorithm~\ref{alg:ICI_approx} and~\ref{alg:ICI_approx heuristic}. As the final remark, the gap in performance of ICZF indicates that a large portion of the optimal resource allocation gain comes from inter-cell coordination.
\begin{figure*}
	\centering
	\begin{subfigure}{.5\textwidth}
		\centering
		\includegraphics[ width=1\columnwidth]{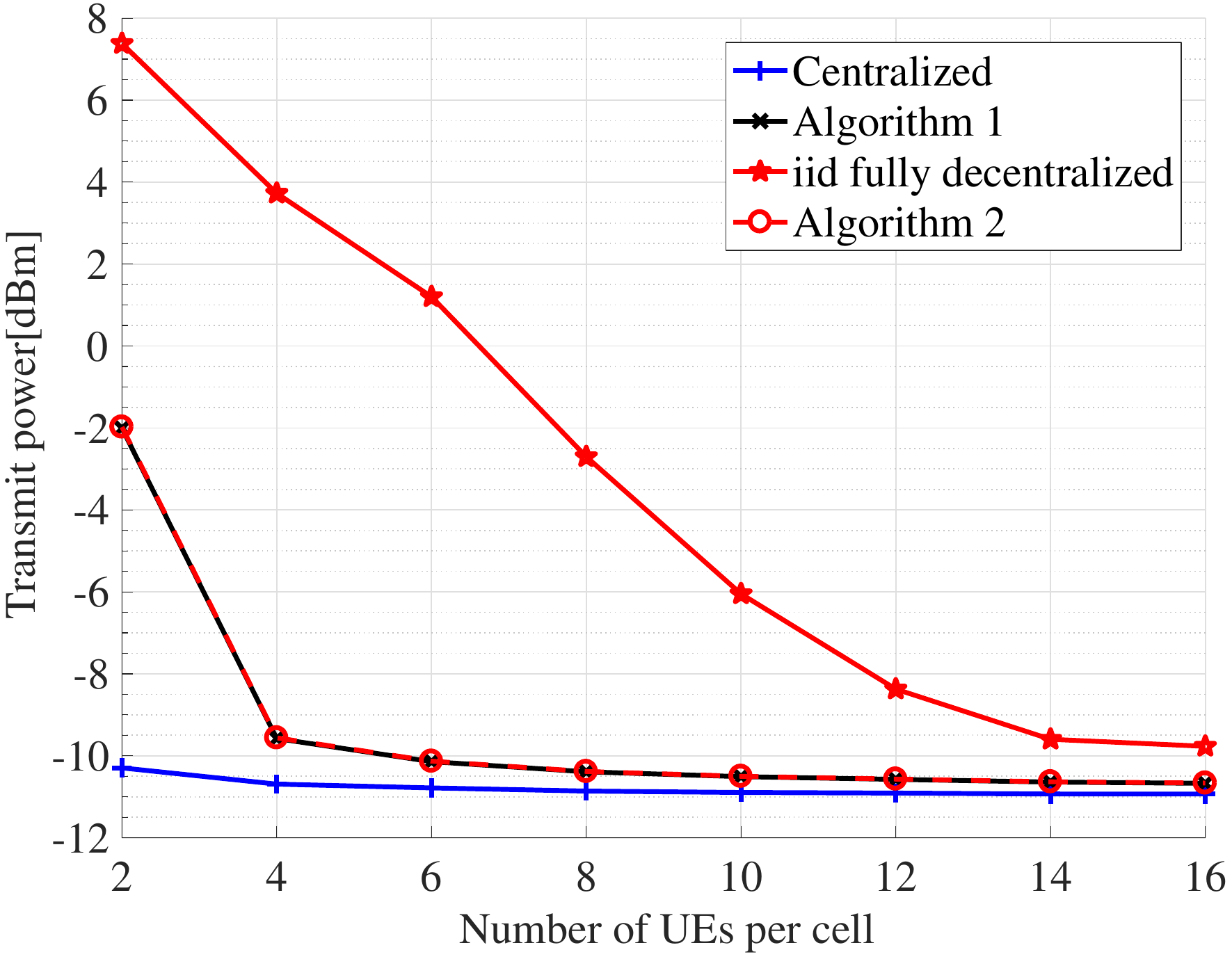}
		\caption{ The scenario with $N=K$.}
	\end{subfigure}%
	\begin{subfigure}{.5\textwidth}
		\centering
		\includegraphics[width=\columnwidth]{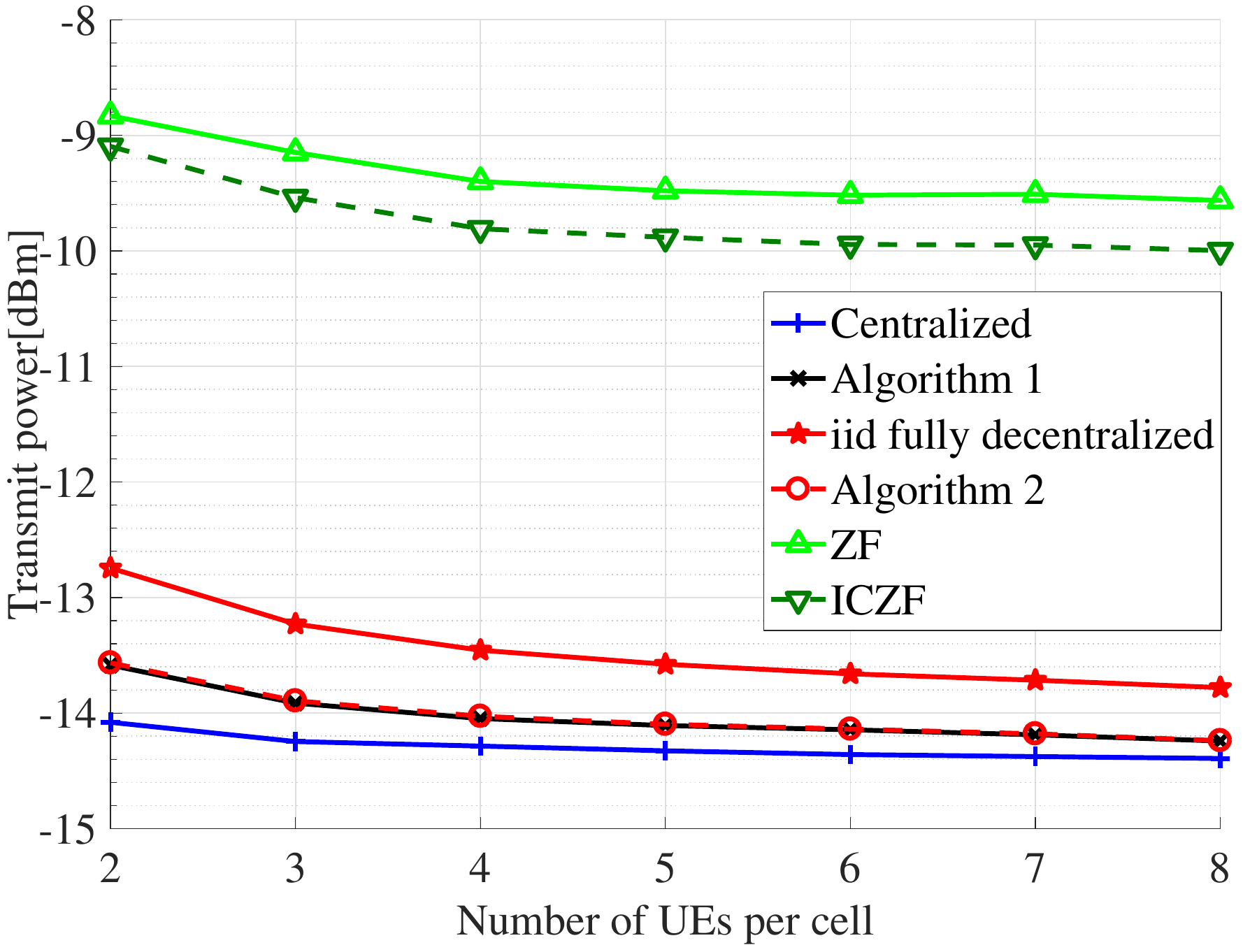}
		\caption{ The scenario with $N=2K$.}
	\end{subfigure}
	\caption{Transmit power vs. $\bar K$, target rate = 1 bits/s/Hz/UE. }
	\label{fig:perfcomp SNR1}
\end{figure*}
%

%

\subsection{The Scenario With Partitioned UE Population }
\begin{figure}[!tbp]
	\centering
	{\includegraphics[width=0.5\columnwidth]{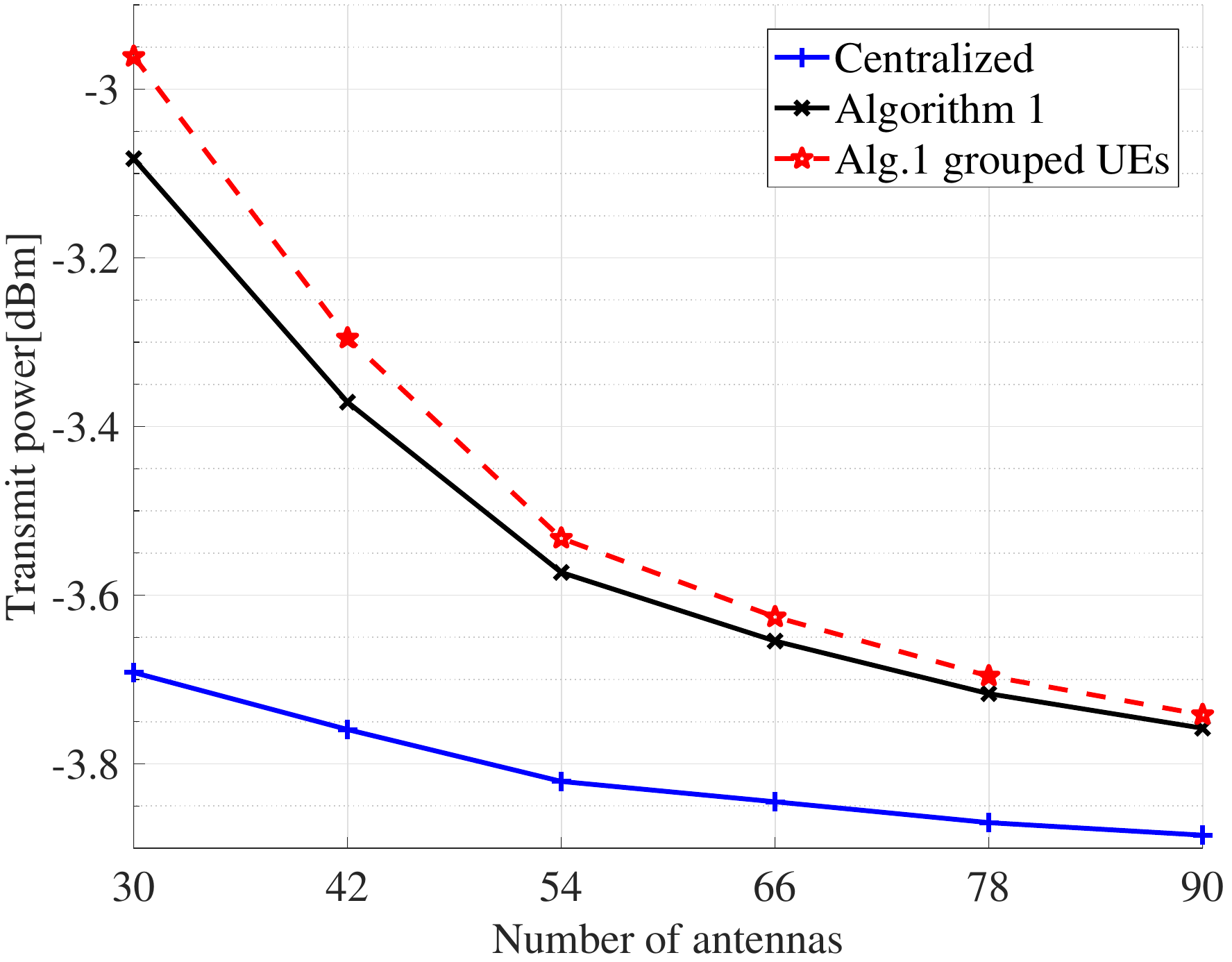}}
	\caption{Transmit power with grouped UEs versus $N$ when $\frac{N}{K}=1$ and the target rate is $1$ bits/s/Hz/UE.}
	\label{fig:GroupUEs Sim Res}
\end{figure}
{In the following, the analysis in Section~\ref{sec:grouped UEs} is validated numerically. We consider a two-cell configuration with the same number of UEs assigned to each cell. Additionally, UEs within a cell are divided equally among three groups according to the arrangement presented in Fig.~\ref{fig:Groups}.
	  We note that the system model considered in Section~\ref{sec:grouped UEs} is generic and can be applied to larger configurations. However, here we reused the two-cell example illustrated in Fig.~\ref{fig:Groups} to avoid redundancy and to keep the numerical results consistent with the explanations in Section~\ref{sec:grouped UEs}.}
 As in Corollary~\ref{cor:homo closeICI}, the UEs within each group are assumed to have an identical correlation matrix but with distinct user specific pathloss values.
The angular spread for all groups are assumed to be equal to $\frac{\pi}{6}$ and the group-specific correlation matrices are evaluated using~\eqref{eq:corr model}. The groups have disjoint angular spreads, and hence, have orthogonal correlation eigenspaces when $N\rightarrow \infty$. However, in finite dimensions, the correlation matrices of distinct groups are not fully orthogonal. The reference curve in Fig.~\ref{fig:GroupUEs Sim Res} titled 'Algorithm 1' calculates approximate ICI values based on~\eqref{3.4} where the non-orthogonality of correlation matrices of distinct groups is considered in the ICI approximations. On the other hand, the curve titled 'Alg.1 grouped UEs' relies on Corollary~\ref{cor:homo closeICI} for deriving approximate ICIs.
In the latter case, the UEs in distinct groups are assumed to have orthogonal correlation matrices.
The precoders in both cases are derived as in Algorithm~\ref{alg:ICI_approx} while utilizing the corresponding ICI approximations. 
We observe a small difference in the transmission power of these two cases, which is due to inter-group orthogonality condition being satisfied only asymptotically. 
The corresponding curves converge and both curves approach the optimal solution as dimensions are increased.

\section{Conclusions and Discussions}
\label{sec:Conclusion} 
In this work, a decentralization framework was proposed for the power minimization problem in multicell MU-MIMO networks. The proposed decentralized solutions attain the QoS-guaranteed precoders locally subject to relaxed coordination requirements. This is particularly important in practice where the backhaul links suffer from imperfections, including limited capacity and latency. The analysis under the assumption of partitioned UE population allowed the ICIs, as the inter-cell coordination messages, to be characterized explicitly in terms of channel statistics. This, in particular, provided insight into the coordination mechanism. Also, it reduced the computational complexity of the approximate ICIs. 
This analysis can be further exploited to attain substantial complexity reduction in the precoding phase as well. This approach generally motivates per-group processing, which is reminiscent of two-stage beamforming~\cite{UserPartionAdhikary}.
To this end, the UE population is first divided into multiple groups each with approximately the same channel correlation matrix. Then, the BSs get approximations of interference leakage in-between groups at a given BS and across cells using results of Theorem~\ref{th:down pw updwn duality} and Corollary~\ref{cor:homo closeICI}. Declaring the group specific interference terms as constraints in the optimization problem, similar to ICI constraints introduced in~\eqref{Opt_problem}, the BSs get the precoders for UEs in a group independently by exploring CSI within each group. The search space, in this case, is limited to the group's degrees of freedom, which is generally much smaller than $N$, and thus results into a significant complexity reduction. 
 
 As a final remark, we notice that the weighting factors $\{\mu_{b}\}$ in~\eqref{eq:prim problemSimple} provide a mechanism
	to tradeoff the power consumption at different BSs~\cite{Dahrouj-Yu-10}. In addition, one can rely on the results of Theorems~\ref{th:up powers updw duality} and~\ref{th:down pw updwn duality} to get a good approximation for total power consumption at BSs based on statistical information of channel vectors. This can be utilized in a feasibility assessment to ensure BS-specific power limits where, as in~\cite{JointAdmisPwrCtrl,JointSchedPWr,PowrCTrlBook}, the infeasibility implies that some UEs should be dropped (admission control methods)~\cite{JointAdmisPwrCtrl} or rescheduled in orthogonal dimensions (scheduling methods)~\cite{JointSchedPWr}. 
 
\appendices
\section{Important theorems and lemmas}
\label{chap:RMT}
In the derivation of large system analysis, we use well-known lemmas including 
trace lemma~\cite[Lemma 2.6]{bai1998no},\cite[Theorem 3.4]{RMT} along with rank-1 perturbation lemma~\cite[Lemma 2.6]{silverstein1995empirical},\cite[Theorem 3.9]{RMT}. The former one shows asymptotic convergence of $\textbf{x}\herm \textbf{A} \textbf{x}-\frac{1}{N}\text{Tr}{\bf{A}}\rightarrow0$ when ${\bf x}\in \mathbb{C}^N$ has i.i.d entries with zero mean, variance of $\frac{1}{N}$ and independent of $\mathbf{A}$. The latter one states that addition of rank-1 matrix ${\bf xx}\herm$ to the random Gram matrix ${\bf{X}} {\bf X}\herm$ does not affect trace $\frac{1}{N}\text{Tr}({\bf X} {\bf X}\herm+{\bf I}_N)$ term in the large dimensional limit. The formal presentation of these lemmas are given in~\cite{bai1998no,silverstein1995empirical,RMT}. 

The other result from random matrix theory that characterizes the so-called Stieltjes transform of the Gram matrix~\cite{RMT} of the propagation channel matrix is given in the following theorem.
\begin{theorem}
	\label{th:main theorem}
	\cite[Theorem 1]{wagner2012large}
	Consider a channel matrix $\mathbf{H} \in \mathbb{C}^{N \times K}$ with columns $ \mathbf{h}_k=\boldsymbol{\Theta}_k^{\frac{1}{2}}  \mathbf{z}_k$, where correlation matrices $\boldsymbol{\Theta}_k =\boldsymbol{\Theta}_k^{\frac{1}{2}}(\boldsymbol{\Theta}_k  \herm )^{\frac{1}{2}} $ are subject to Assumption~\ref{as:1} and vectors ${\vec z}_j \in \mathbb{C}^N$ have zero mean i.i.d entries of variance $\frac{1}{N}$ and eighth-order moment of order $\mathcal{O}(\frac{1}{N^4})$. Then, for $\mathbf{H} \bf{C} \mathbf{H}^{\mathrm{\scriptsize H}}$ with $\bf{C}=\rm{diag}\{c_1,...,c_K\} $ where $c_k,\,\forall k$ are finite deterministic values, we define
	\begin{equation}\label{theroma1}
	m_{k,i }(z,x)\triangleq \frac{1}{N}{\rm Tr}\biggr(\boldsymbol{\Theta}_k(\mathbf{H} {\bf{C}}  \mathbf{H}^{\mathrm{\scriptsize H}}+x\boldsymbol{\Theta}_i-z\mathbf{I}_N)^{-1}\biggr)
	\end{equation}
	where for $z \in \mathbb{C} \backslash \mathbb{R}^+$ and bounded positive variable $x$, when dimensions $K$ and $N$ grows large with fixed ratio of $\frac{N}{K} < \infty$, it follows that
 	$m_{k,i }(z,x)-\bar{m}_{k,i }(z,x)\xrightarrow{N\rightarrow \infty} 0$
	almost surely and the deterministic equivalent is given by
	\begin{equation}\label{Proof equivalent ST form}
	\begin{aligned}
	\bar{m}_{k,i}(z,x)=\frac{1}{N} \!{\rm Tr }\biggr(\boldsymbol{\Theta}_k\biggr(\frac{1}{N}\sum\limits_{j=1}^K \frac{c_j \boldsymbol{\Theta}_j}{1+c_j\bar{m}_{j,i}(z,x)}+x\boldsymbol{\Theta}_i-z\mathbf{I}_N\biggr)^{-1}\biggr).
	\end{aligned}
	\end{equation}
\end{theorem}
\begin{IEEEproof} The fundamental idea of the proof is based on Bai and Silverstein technique~\cite{RMT} where the deterministic equivalent of $m_{k,i }(z,x)$ is inferred by writing it under the form $\frac{1}{N}\text{Tr}({\bf D}^{-1})$ where $\bf D$ needs to be determined. This is performed by selecting ${\bf{D}}=(\mathbf{R}+x\boldsymbol{\Theta}_i-z\mathbf{I}_N)\boldsymbol{\Theta}_k^{-1}$ and then taking the difference $m_{k,i }(z,x)-\frac{1}{N}\text{Tr}({\bf D}^{-1})\rightarrow0$. Then, utilizing random matrix theory results, the deterministic matrix $\bf R$ is determined such that the difference tends to zero almost surely.
The formal proof of the theorem in a more generic configuration is given in~\cite{wagner2012large}.\end{IEEEproof}

	For clarity, we simplify $m_{k,i}(z,x)$ and $\bar{m}_{k,i}(z,x)$ notations to reflect specific settings. In particular, we drop the index $i$ and variable $x$ in $\bar{ m}_{k,i}(z,x)$ in the cases with $x=0$, i.e., $\bar{m}_{k}(z)=\bar{m}_{k,i}(z,0)$. Also,  when $z$ is equal to the noise variance, we simply refer to $\bar{m}_{k}(z)$ by $\bar{m}_{k}$. In multicell scenario the measures carry a BS index as well, for example, $\bar{m}_{b.k}$ refers to the measure corresponding to the channel matrix at BS $b$. 
	
\section{Lagrangian Duality Analysis}	
\label{sec:duality analysis}
 		In this appendix, we formulate the {Lagrange dual problem} of~\eqref{Opt_problem}. To do so, we first show that the {Lagrange dual problem} of~\eqref{Opt_problem} is the same as that of~\eqref{eq:prim problemSimple}. To this end, we first write the Lagrangian of~\eqref{Opt_problem} as  
 		\begin{equation}
 		 			\begin{aligned}\nonumber
		L({\bf w}_{k},\lambda_k,\beta_{b,k},\epsilon_{b,k})  &=\sum_{k=1}^{K}  \mu_{b_k}\left\|{\bf w}_{k} \right\|^{2}-
		\sum_{k=1}^{K} \frac{\lambda_k}{N} \Biggr(\frac{\left|{\bf h}_{b_k,k}\herm	{\bf w}_{k}\right|^{2}}{\gamma_k}-\!\!\!\sum\limits_{i\in {\mathcal U}_{b_{k}}\setminus k} \left|{\bf h}_{b_k,k}\herm{\bf w}_{i}\right|^{2}-\!\! \sum\limits_{b\in{\mathcal B}\setminus b_{k}} \epsilon_{b,k} -  \sigma^{2} \Biggr)\\
		& + \sum\limits_{b}\sum\limits_{k\notin \mathcal{U}_{b}}\beta_{b,k} \Biggr(\sum\limits_{j\in \mathcal{U}_{b}} \left|{\bf h}_{b,k}\herm{\bf w}_{j}\right|^{2}-\epsilon_{b,k}\Biggr)
		\end{aligned}
			\end{equation}
		where $\{\lambda_k\}$ and $\{\beta_{b,k}\}$ are the Lagrange dual variables associated with SINR and ICI constraints, respectively. From complementary slackness~\cite{Boyd-Vandenberghe-04}, we know that either $\beta_{b,k}$ or $\sum_{j\in \mathcal{U}_{b}} \left|{\bf h}_{b,k}\herm{\bf w}_{j}\right|^{2}-\epsilon_{b,k}$ have to be equal to zero. By setting the variable $\beta_{b,k}$ to zero, the variable $\epsilon_{b,k}$ becomes unconstrained, and thus, can be chosen to make
		$	{\text{min}}_{{\bf w}_{k},\epsilon_{b,k},\lambda_k} \,\, L({\bf w}_{k},\lambda_k,\beta_{b,k},\epsilon_{b,k})=- \infty
		$
		. This suggests that $\beta_{b,k}, \forall b,k$ have non-zero values and consequently complementary slackness implies the equality $\epsilon_{b,k}=\sum_{j\in \mathcal{U}_{b}} \left|{\bf h}_{b,k}\herm{\bf w}_{j}\right|^{2}$ to hold. By plugging this into the Lagrangian, we can follow the same approach as in~\cite{Dahrouj-Yu-10} to obtain the {Lagrange dual problem} of~\eqref{eq:prim problemSimple} and~\eqref{Opt_problem} as\footnote{{The Lagrangian duality between~\eqref{eq:prim problemSimple} and~\eqref{eq:dual uplink} holds when both~\eqref{eq:prim problemSimple} and~\eqref{eq:dual uplink}  are feasible.}}
		\begin{equation}\label{eq:dual uplink}
		\begin{aligned}
		&\underset{{{\{{\bf v}_{k}\}}} , \{\lambda_{k}\}}{\min}
		\quad \sum_{b{\in}\mathcal{B}}\sum_{k{\in}\mathcal{U}_{b}} \frac{\lambda_{k}}{N}\sigma^2 \\
		&\quad\text{s.t.}\quad   \frac{\lambda_{k}|{{\vec v}}_{k}\herm{\vec h}_{b_{k},k}|^2}
		{
			\sum_{j\in \mathcal U \setminus k} \lambda_{j} |{{\vec v}}_{k} \herm{\vec h}_{b_{k},j}|^2+ \mu_{{b_k}}N\|{{\vec v}}_{k}\|^2
		}  \geq {\gamma_{k}}, \;\forall k \in \mathcal{U},
		\end{aligned}
		\end{equation}
		where the Lagrange dual variables ${\lambda_k}/{N}$  can be thought of as the UE power in the dual uplink power minimization problem~\cite{ Dahrouj-Yu-10}. The optimal receive beamforming vectors $\{{\vec v}_{k}\}$ are given as a set of minimum mean square error (MMSE) receivers $
		\mathbf{v}_{k}=(\sum_{j\in \mathcal U\setminus k}{\lambda_{j}}\mathbf{  h}_{b_{k},j}\mathbf{  h}_{b_{k},j}\herm + \mu_{b_k}N\mathbf{I}_{N})^{-1}\mathbf{  h}_{b_k,k}$, and the optimal Lagrangian multipliers $\boldsymbol{\lambda}^*= [\lambda_{1}^*, \ldots, \lambda_{K}^*]\tran $ are obtained as in~\eqref{eq:lambda itr}.
		The duality condition provides the downlink precoders as ${{\vec w}}_{k}=\sqrt{{{\delta_{k}}}/{N}} {\vec v}_{k}$ with the scaling factors $\{\delta_{k}\}$ given as $\boldsymbol{\delta} =  \mathbf{G}^{-1} \mathbf{1}_{K}\sigma^2$ where $\boldsymbol{\delta} = [\delta_{1}, \ldots, \delta_{K}]\tran$, and the $(i,k)^{\text{th}}$ element of the so-called coupling matrix $\mathbf{G} \in \mathbb C^{K\times K}$ is given as in~\eqref{eq:G_matrix}.
\section{Proof of Theorem \ref{th:up powers updw duality}}
\label{sec:proof Theorem1}
Without loss of generality, the noise power $\sigma^2$ and BS power weights $\mu_{b},\,\forall b$ are assumed to be equal to one in the following analyses. In order to prove the theorem, we first provide an intuition by heuristically assuming the Lagrangian multipliers to be deterministic values and given independently of channel vectors. Then we prove the convergence of the optimal Lagrangian multipliers in a feasible problem to the deterministic equivalents rigorously via a contradiction argument.
Denoting $\boldsymbol{\lambda}^*$ as the fixed point solution to~\eqref{eq:lambda itr} in a feasible optimization problem, the following equality holds
\begin{align}
\label{eq:lambda itr proofs}
\frac{\gamma_{k}}{\lambda_{k}^*}= \frac{1}{N}{\mathbf{h}_{b_k,k}\herm\left(\sum\limits_{j\in \mathcal U\setminus k}\frac{\lambda_{j}^*}{N}\mathbf{  h}_{b_{k},j}\mathbf{  h}_{b_{k},j}\herm + \mathbf{I}_{N}\right)^{-1}\mathbf{h}_{b_k,k}}
\end{align}
where the superscript $^*$ stands for the optimal solution. Assuming erroneously that $\boldsymbol{\lambda}^*$ is given and independent of channel vectors, trace lemma~\cite[Lemma 2.6]{bai1998no} along with rank-1 perturbation lemma~\cite[Lemma 2.6]{silverstein1995empirical} in~\eqref{eq:lambda itr proofs} yields the following
$\frac{\gamma_{k}}{\lambda_{k}^*}-\frac{1}{N} {\rm{Tr}}\big(\boldsymbol{\Theta}_{b_{k},k} \big(\sum_{j\in \mathcal U}\frac{\lambda_{j}^*}{N}\mathbf{  h}_{b_{k},j}\mathbf{  h}_{b_{k},j}\herm+\mathbf{I}_N \big)^{-1}\big) \rightarrow 0$, almost surely. This trace term is equivalent to $m_{b_k,k}^*$ in Theorem~\ref{th:main theorem} that satisfies an almost sure convergence $m_{b_k,k}^*-\bar{m}_{b_k,k}^*\rightarrow 0$ where $\bar{m}_{b_k,k}^*$ is given as a solution of a system of equations
\vspace{-0.1cm}
\begin{align}
	\bar{m}_{b_k,k}^* = \frac{1}{N}{\rm {Tr}}\biggr({\bf \Theta}_{b_k,k}\bigg(\frac{1}{N}\sum_{j\in \mathcal U}\frac{ \lambda_{j}^* {\bf \Theta}_{b_k,j}}{1+\lambda_j^*\bar{m}_{b_k,j}^*} + {\bf I}_{N}\bigg)^{-1}\biggr).
\end{align}
From the above discussion, we may then expect the terms $\{{\lambda}_k^*\}$ to be all close to $\{\gamma_k/\bar{m}_{b_k,k}^*\}$ for $N$ and $K$ large enough.
However, since the optimal Lagrangian multipliers depend on the channel vectors, we cannot rely on classical random matrix theory results for proving the asymptotic convergence of $\boldsymbol{\lambda}^*$ to the deterministic equivalents. Thus, we follow the approach introduced in~\cite{couillet2014large,LucaCouilletDebbahJournal2015}, and prove the asymptotic convergence of $\boldsymbol{\lambda}^*$ via a contradiction argument. In particular, we 
set $\bar \lambda_k=\gamma_k/\bar{m}_{b_k,k}$ with $\bar{m}_{b_k,k}$ given as
\begin{align}
\label{eq:stform proof of th1}
\bar{m}_{b_k,k} = \frac{1}{N}{\rm {Tr}}\biggr({\bf \Theta}_{b_k,k}\biggr(\frac{1}{N}\sum_{j\in \mathcal U}\frac{\bar \lambda_{j} {\bf \Theta}_{b_k,j}}{1+\bar\lambda_j\bar{m}_{b_k,j}} + {\bf I}_{N}\biggr)^{-1}\biggr).
\end{align}
Then we show via a contradiction argument that the ratios $r_k=\frac{\bar \lambda_k}{\lambda_k^*}=\frac{ \gamma_k}{\bar{m}_{b_k,k}}\frac{1}{\lambda_k^*},\,\forall k\in \mathcal{U}$ converge asymptotically to one, which allows the results of the theorem to be claimed. 
To do so, we consider BS $b$ with the set of served UEs $\mathcal{U}_b=\{\rm{UE}_1,...,\rm{UE}_M\}$ where, given the ratios $r_k,\,\forall k\in \mathcal{U}_b$, equation~\eqref{eq:lambda itr proofs} can be rewritten as
\begin{align}
\label{eq:rewrite as ratio}
\frac{r_k\gamma_{k}}{\bar\lambda_{k}}= \frac{1}{N}{{\bf z}_{b,k}\herm{\bf \Theta}_{b,k}^{1/2}\left(\sum\limits_{j\in \mathcal U\setminus k}\frac{\bar \lambda_{j}}{r_j}{\bf B}_j+ \mathbf{I}_{N}\right)^{-1}{\bf \Theta}_{b,k}^{1/2}{\bf z}_{b,k}},
\end{align}
where ${\bf B}_j=\frac{1}{{N}}{\bf \Theta}_{b,j}^{1/2}{{\bf z}_{b,j}{\bf z}_{b,j}\herm}{\bf \Theta}_{b,j}^{1/2}$. Next, the UE indexes in $\mathcal{U}_b$ are relabeled such that the following holds $0\leq r_1 \leq...\leq r_M$
with $\{r_j\}$ assumed to be well defined and positive. Rewriting~\eqref{eq:rewrite as ratio} for UE $M\in\mathcal{U}_b$ and replacing all $r_j,\forall j\in \mathcal{U}$ in the summation with the largest ratio $r_M$, 
 we get the following inequality, based on monotonicity
 arguments,
\begin{align}
\label{eq:inequalality0}
\frac{r_M\gamma_{M}}{\bar\lambda_{M}}\leq \frac{1}{N} {{\bf{z}}_{b,M}\herm  {\boldsymbol\Theta}_{b,M}^{1/2}\left(\sum\limits_{j\in \mathcal U\setminus M}\frac{\bar \lambda_{j}}{r_M}{\bf B}_j+ \mathbf{I}_{N}\right)^{-1}\!\!\!\!{\bf \Theta}_{b,M}^{1/2} {\bf{z}}_{b,M}},
\end{align}
or equivalently
\begin{align}
\label{eq:inequality1}
\frac{\gamma_{M}}{\bar\lambda_{M}}\leq \frac{1}{N}{{\bf{z}}_{b,M}^{\mathrm{\scriptsize H}}  {\boldsymbol\Theta}_{b,M}^{1/2}\left(\sum\limits_{j\in \mathcal U\setminus M}{\bar \lambda_{j}}
{\bf B}_j
	+ {r_M}\mathbf{I}_{N}\right)^{-1}{\!\!\!\!\bf \Theta}_{b,M}^{1/2} {\bf{z}}_{b,M}}.
\end{align}
Assume now that $r_M$ is infinitely often larger than $1+l$ with $l> 0$ some positive value. Restricting ourselves to such a subsequence, the monotonicity arguments give the inequality in~\eqref{eq:inequality1} equivalently as
\begin{align}
\label{eq:inequality2}
\!\!\!\frac{\gamma_{M}}{\bar\lambda_{M}}\leq\frac{1}{N}{{\bf{z}}_{b,M}\herm  {\boldsymbol\Theta}_{b,M}^{1/2}\left(\sum\limits_{j\in \mathcal U\setminus M}{\bar \lambda_{j}}
{\bf B}_j
	+ {(1+l)}\mathbf{I}_{N}\right)^{-1}\!\!\!\!\!\!{\bf \Theta}_{b,M}^{1/2}{\bf{z}}_{b,M}}.
\end{align}
Denoting the right hand side of~\eqref{eq:inequality2} by $\iota _M$, we observe that  $\iota _M$ does not depend anymore on $\boldsymbol{\lambda}^*$. Thus, we  can apply trace lemma~\cite[Lemma 2.6]{bai1998no}, rank-1 perturbation lemma~\cite[Lemma 2.6]{silverstein1995empirical} and Theorem~\ref{th:main theorem} to the right hand side of the above inequality to get $
\iota _M-\bar{m}_{b_M,M}(-(1+l))\xrightarrow{N\rightarrow \infty} 0
$ with
\begin{align}
\bar{m}_{b_M,M}(z) = {\rm {Tr}}\biggr({\bf \Theta}_{b_M,M}\biggr(\sum_{j\in \mathcal U}\frac{\bar \lambda_{j} {\bf \Theta}_{b_M,j}}{1+\bar\lambda_j\bar{m}_{b_M,j}}\!-zN {\bf I}_{N}\biggr)^{-1}\biggr),
\end{align}
which along with~\eqref{eq:inequality2} results in $\frac{\gamma_{M}}{\bar\lambda_{M}}\leq\lim_{N \to \infty} \text{inf}\,\, \bar{m}_{b_M,M}(-(1+l))$.
On the other hand, we notice that $\bar{m}_{b_M,M}(-(1+l))$ at $l=0$ is equal to $\bar{m}_{b_M,M}$ in~\eqref{eq:stform proof of th1} with $\bar \lambda_M=\gamma_M/\bar{m}_{b_M,M}$. Since $\bar{m}_{b_M,M}(-(1+l))$ is a decreasing function of $l$ it can be proved~\cite{couillet2014large} that for any $l>0$ we have $\lim_{N \to \infty} \text{sup}\,\, \bar{m}_{b_M,M}(-(1+l))<\frac{\gamma_{M}}{\bar\lambda_{M}}$.
This, however, goes against the former condition and creates a contradiction on the initial hypothesis that $r_M> 1+l$ infinitely often. Therefore, we must admit that $r_M\leq 1+l$ for all large values of $N$ and $K$. Reverting all inequalities and using similar arguments yields $r_1\geq 1-l$  for all large values of $N$ and $K$.  
Putting all these results together yields $1-l\leq r_1,...,r_M\leq 1+l$,  from
which we may write ${\rm{max}}_{k\in \mathcal{U}_b}|r_k-1|\leq l$ for all large values of $N$ and $K$.
Taking a countable
sequence of $l$ going to zero, we eventually obtain ${\rm{max}}_{k\in \mathcal{U}_b}|r_k-1|\rightarrow0$. Noticing that $r_k=\frac{ \gamma_k}{\bar{m}_{b_k,k}\lambda_k^*}$, we get ${\rm{max}}_{k\in \mathcal{U}_b}|\lambda_k^*-\bar \lambda_k|\rightarrow 0$ almost surely with $\bar \lambda_k=\frac{ \gamma_k}{\bar{m}_{b_k,k}}$. Following the same steps for all other $b\in \mathcal{B}$ completes the proof.

\section{Proof of Theorem \ref{th:down pw updwn duality}}
\label{sec:proof Theorem2}
Given the Lagrangian multipliers $\{\bar{{\lambda}}_k\}$ as deterministic values in a feasible problem from Theorem~\ref{th:up powers updw duality}, we derive the deterministic equivalents for the elements of coupling matrix in~\eqref{eq:G_matrix} using standard random matrix theory tools. To do so, we rewrite the diagonal elements of the coupling matrix as $[\mathbf{G}]_{k,k}=\frac{1}{\gamma_{k}}{|\frac{1}{N}{\mathbf{h}_{b_k,k}\herm\boldsymbol{\Sigma}_{b_{k}}^{\backslash k}\,\mathbf{h}_{b_k,k}}|^2}$ where   $\boldsymbol{\Sigma}_{b_{k}}^{\backslash k}=\big(\sum_{j\in \mathcal U\setminus k}\frac{\bar{\lambda}_{j}}{N}\mathbf{  h}_{b_{k},j}\mathbf{  h}_{b_{k},j}\herm  +\mathbf{I}_N\big)^{-1}$ and the notation $()^{\backslash k}$ excludes the $k^{\text{th}}$ term from the summation.  Given the growth rate in Assumption~\ref{as:0}, we apply trace lemma~\cite[Lemma 2.6]{bai1998no} and rank-1 perturbation lemma~\cite[Lemma 2.6]{silverstein1995empirical} which, gives $\frac{1}{N}{\mathbf{h}_{b_k,k}\herm\boldsymbol{\Sigma}_{b_{k}}^{\backslash k}\,\mathbf{h}_{b_k,k}}-\frac{1}{N} \text{Tr}(\boldsymbol{\Theta}_{b_{k},k} \boldsymbol{\Sigma}_{b_{k}})\rightarrow 0 $ almost surely. The resulted trace term is equal to $m_{b_k,k}$ defined in Theorem~\ref{th:main theorem} where, according to the theorem, we have $m_{b_k,k}-\bar{m}_{b_k,k}\rightarrow 0$ almost surely. This implies $[\mathbf{G}]_{k,k}-\frac{1}{\gamma_k}\bar{m}_{b_k,k}^2\rightarrow 0$ almost surely, which gives the diagonal elements of the coupling matrix as stated in the theorem. 

The non-diagonal elements of the coupling matrix $
[\mathbf{G}]_{k,i}=-\frac{1}{N^2}{\vec h}_{b_{i},k}\herm \boldsymbol{\Sigma}^{\backslash i}_{b_{i}} \,{ {\vec h}_{b_{i},i} {\vec h}_{b_{i},i}\herm }\boldsymbol{\Sigma}^{\backslash i}_{b_{i}}  \, {\vec h}_{b_{i},k}$ can be rewritten using matrix inversion lemma~\cite[Equation 2.2]{silverstein1995empirical} as
\begin{equation}\label{eq:Gil l remov}
[\mathbf{G}]_{k,i}=-\frac{1}{N^2}\frac{{\vec h}_{b_{i},k}\herm \boldsymbol{\Sigma}^{\backslash i,k}_{b_{i}} \,  {\vec h}_{b_{i},i} {\vec h}_{b_{i},i}\herm \boldsymbol{\Sigma}^{\backslash i,k}_{b_{i}}\, {\vec h}_{b_{i},k}}{\big(1+\frac{\bar \lambda_k}{N}{\vec h}_{b_{i},k}\herm \boldsymbol{\Sigma}^{\backslash i,k}_{b_{i}} \, {\vec h}_{b_{i},k}\big)^2},
\end{equation}
where  $\boldsymbol{\Sigma}_{b_{k}}^{\backslash i, k}=\big(\sum_{j\in \mathcal U\setminus i, k}\frac{\bar{\lambda}_{j}}{N}\mathbf{  h}_{b_{k},j}\mathbf{  h}_{b_{k},j}\herm  +\mathbf{I}_N\big)^{-1}$ with notation $()^{\backslash i, k}$ excluding the $i^\text{th}$ and $k^{\text{th}}$ terms from the summation. Now, we can apply trace lemma~\cite[Lemma 2.6]{bai1998no} and rank-1 perturbation lemma~\cite[Lemma 2.6]{silverstein1995empirical} to the denominator of~\eqref{eq:Gil l remov} to obtain $\frac{1}{N}{{\vec h}_{b_{i},k}\herm \boldsymbol{\Sigma}^{\backslash i,k}_{b_{i}}\, {\vec h}_{b_{i},k}}-{\frac{1}{N}\text{Tr} (\boldsymbol{\Theta}_{b_i,k} \boldsymbol{\Sigma}_{b_{i}} })\, \rightarrow 0$ almost surely. Therefore, as a result of Theorem~\ref{th:main theorem}, we get the almost sure convergence of the denominator as
\begin{equation}
\label{eq:denom G}
\bigg(1+\frac{\bar \lambda_k}{N}{\vec h}_{b_{i},k}\herm \boldsymbol{\Sigma}^{\backslash i,k}_{b_{i}} \, {\vec h}_{b_{i},k}\bigg)^2-\left(1+\bar \lambda_k \bar m_{b_i,k}\right)^2\rightarrow 0.
\end{equation}
We proceed by applying trace lemma~\cite[Lemma 2.6]{bai1998no} to the numerator of~\eqref{eq:Gil l remov} that gives
\begin{equation}\label{eq:Gli tr}
\begin{aligned}
\hspace{-1.5cm}\frac{1}{N^2}{\vec h}_{b_{i},k}\herm \boldsymbol{\Sigma}^{\backslash i,k}_{b_{i}}\, {{\vec h}_{b_{i},i} {\vec h}_{b_{i},i}\herm} \boldsymbol{\Sigma}^{\backslash i,k}_{b_{i}} \,{\vec h}_{b_{i},k} - 
\frac{1}{N^2} \text{Tr}({\boldsymbol{\Theta} }_{b_{i},k} \boldsymbol{\Sigma}^{\backslash i,k}_{b_{i}}\,  {{\vec h}_{b_{i},i} {\vec h}_{b_{i},i}\herm} \boldsymbol{\Sigma}^{\backslash i,k}_{b_{i}})\rightarrow 0
\end{aligned}
\end{equation}
almost surely. 
 Rearranging the terms inside the trace in~\eqref{eq:Gli tr} and reapplying trace lemma~\cite[Lemma 2.6]{bai1998no} yields 
  \begin{equation}\label{eg:numerator}
{\vec h}_{b_{i},i} \herm \boldsymbol{\Sigma}^{\backslash i,k}_{b_{i}} 
\frac{{\boldsymbol{\Theta} }_{b_{i},k}}{N^2}
\boldsymbol{\Sigma}^{\backslash i,k}_{b_{i}} \, {\vec h}_{b_{i},i} -
 \text{Tr}  \big(\boldsymbol{\Theta}_{b_{i},i} \boldsymbol{\Sigma}_{b_{i}}^{\backslash i,k} 
\frac{\boldsymbol{\Theta}_{b_{i},k}}{N^2}
\boldsymbol{\Sigma}_{b_{i}}^{ \backslash i,k}\big) \rightarrow 0
 \end{equation}
almost surely. As a result of rank-1 perturbation lemma~\cite[Lemma 2.6]{silverstein1995empirical}, the $i^\text{th}$ and $k^\text{th}$  excluded terms in $\boldsymbol{\Sigma}_{b_{i}}^{ \backslash i,k}$ are asymptotically insignificant in the trace, and thus,~\eqref{eq:denom G}-\eqref{eg:numerator} give the almost sure convergence of~\eqref{eq:Gil l remov} as
\begin{equation}\label{eq:Glk determins meanway}
\hspace{-0.1 in}(-[\mathbf{G}]_{k,i})-\frac{\frac{1}{N^2} \text{Tr}(  \boldsymbol{\Theta}_{b_{i},i} \boldsymbol{\Sigma}_{b_{i}} \,
	\boldsymbol{\Theta}_{b_{i},k}
	\boldsymbol{\Sigma}_{b_{i}})}{\left(1+\bar \lambda_k \bar m_{b_i,k}\right)^2}\rightarrow 0.
\end{equation}
From matrix identities~\cite{Horn-Johnson-90}, we know that $ 
{\partial \mathbf{Y}^{-1}}/{\partial x}=- \mathbf{Y}^{-1} ( {\partial \mathbf{Y}}/{\partial x} ) \mathbf{Y}^{-1}
$  with $\bf Y$ being a matrix depending on variable $x$. Keeping this in mind, we refer to $m_{b_i,i,k}(z,x)$ in its general form defined in Theorem~\ref{th:main theorem} as $m_{b_i,i,k}(z,x)=\frac{1}{N} \text{Tr}\big(\boldsymbol{\Theta}_{b_{i},i}\big( \sum_{j\in \mathcal U}\frac{\bar{\lambda}_{j}}{N}\mathbf{  h}_{b_{i},j}\mathbf{  h}_{b_{i},j}\herm -z\mathbf{I}_N-x  \boldsymbol{\Theta}_{b_{i},k} \big)^{-1} \big)$ where in the special setting with $x=0$ and $z=-1$ diminishes to $m_{b_i,i}$. Using the identity, the numerator of~\eqref{eq:Glk determins meanway} can be written as a derivative of $m_{b_i,i,k}(z,x)$ with respect to the auxiliary variable $x$ at point $(z=-1,x=0)$, i.e., 
\begin{equation}
\label{eq:numer G}
\frac{1}{N} \text{Tr} \big(\boldsymbol{\Theta}_{b_{i},i} \boldsymbol{\Sigma}_{b_{i}} \,
\boldsymbol{\Theta}_{b_{i},k}
\boldsymbol{\Sigma}_{b_{i}}\big) =\frac{\partial}{\partial x} m_{b_i,i,k}(z,x)|_{x=0,z=-1}.
\end{equation}
Therefore, given the deterministic equivalents of derivative terms $m_{b_i,i,k}'=\frac{\partial}{\partial x} m_{b_i,i,k}(z,x)|_{x=0,z=-1}$ in~\eqref{eq:numer G}, the deterministic equivalents for the non-diagonal elements of the coupling matrix will follow from~\eqref{eq:Glk determins meanway}.
In doing so, we notice that Theorem~\ref{th:main theorem} ensures the almost sure convergence of $m_{b_i,i,k}(z,x)$ to its deterministic equivalent given by $\bar{m}_{b_i,i,k}(z,x)=\frac{1}{N}\text{Tr}(\boldsymbol{\Theta}_{b_i,i}{\mathbf{T}_{b_i,k}(z,x)})$ where
 \begin{equation}\label{eq:Tbkl}
{\mathbf{T}_{b,k}(z,x)}={\bigg(\frac{1}{N} \sum_{j \in \mathcal{U}} \frac{\bar{\lambda}_j \boldsymbol{\Theta}_{b,j}}{1+
	\bar{ \lambda}_j\bar{m}_{b,j,k}(z,x)
} -x\boldsymbol{\Theta}_{b,k}-z \mathbf{I}_{N} \bigg)^{-1}}\!\!\!.
 \end{equation}
Therefore, the deterministic equivalents for the derivative terms, hereafter denoted by $\bar{m}_{b_i,i,k}'$, can be evaluated by deriving the derivative of~\eqref{eq:Tbkl} with respect to $x$ as
\begin{equation}\label{eq:Tbkl prime}
\mathbf{T}_{b,k}^{\prime} = \mathbf{T}_{b}\bigg(\frac{1}{N} \sum_{j \in \mathcal{U}} \frac{\bar{\lambda}_j^2 \boldsymbol{\Theta}_{b,j}\bar{m}_{b,j,k}^{\prime}}{(1+\bar{ \lambda}_j\bar{ m}_{b,j})^2} +\boldsymbol{\Theta}_{b,k} \bigg) \mathbf{T}_{b}
\end{equation}
where $\mathbf{T}_{b}=\mathbf{T}_{b,k}(-1,0)$, $\mathbf{T}_{b,k}^{\prime}=\frac{\partial}{\partial x}\mathbf{T}_{b,k}(-1,x)|_{x=0}$ and $\bar{ m}_{b,j}=\bar{ m}_{b,j,k}(-1,0)$ . Since $\bar{m}_{b_i,i,k}'=\frac{1}{N} \text{Tr} (\boldsymbol{\Theta}_{b_{i},i} \mathbf{T}_{b_i,k}^{\prime})$ with $\mathbf{T}_{b,k}^{\prime}$ given by~\eqref{eq:Tbkl prime}, we get a system of equation to evaluate $\bar{m}_{b_i,i,k}'$ as $  [\bar{m}_{b,1,k}^{\prime},...,\bar{m}_{b,K,k}^{\prime}]$ $=(\mathbf{I}_K-\mathbf{L}_{b})^{-1} {\vec u}_{b,k}$ with ${\vec u}_{b,k}$ and $\mathbf{L}_{b}$ defined as in~\eqref{eq:Proof en prime sys of equations3} and \eqref{eq:Proof en prime sys of equations4}, respectively. Given $\bar{m}_{b_i,i,k}'$, we get the deterministic equivalents for non-diagonal elements of the coupling matrix from~\eqref{eq:Glk determins meanway} and~\eqref{eq:numer G} as $[\mathbf{G}]_{k,i}=-{\frac{1}{N} \bar{m}_{b_i,i,k}'}/{\left(1+\bar \lambda_k \bar m_{b_i,k}\right)^2}$, which completes the proof.

\section{Proof of Corollary \ref{cor:homo closeICI}}
\label{subse:proof of corr ICI}
We notice that the ICI from BS $b$ to UE $k$ in term of downlink transmit powers is given by  $\epsilon_{b,k}=-\sum_{j\in \mathcal U_{b}} p_{j} [{\mathbf{G}}]_{k,j}/\|{\bf v}_j\|^2$ that carries a normalization term compared to the formulation in~\eqref{3.1}. Assuming UE $k$ to belong to a group $g\in \mathcal{A}_b$ with $\mathcal{A}_b$ denoting the set of all groups of BS $b$, one can observe from~\eqref{eq:nonnormal G} that $[\bar{\mathbf{G}}]_{k,j},\,\forall j\not\in \mathcal{G}_g$ is zero (due to inter-group orthogonality assumption) and we get
\begin{equation}\label{eq:epsil group exact}
\bar{\epsilon}_{b,k}=-\sum_{j\in \mathcal U_{b}\cap \mathcal{G}_g } \bar{p}_{j} [\bar{{\mathbf{G}}}]_{k,j}/\|\bar{{\bf v}}_j\|^2
\end{equation} 
where the elements of coupling matrix $\{[\bar{{\mathbf{G}}}]_{k,j}\}$ are given in~\eqref{eq:nonnormal G} as a function of $\bar{m}_{b_j,j,k}'$ and $\bar{m}_{b_j,k}$. The term $\bar{m}_{b_j,j,k}'$ is derivative of $\bar{m}_{b_j,j,k}(z.x)$ with respect to $x$ at point $x=0$ and $z=-1$, and $\bar{m}_{b_j,j,k}(z.x)$ is the Stieltjes transform in its general form as defined in Theorem~\ref{th:main theorem}.
Given identical correlation properties for UEs within a group we can introduce group specific parameters 
 $\bar{\eta}_{b,g}=\bar{m}_{b,j}/a_{b,j}^2,\,\forall j\in \mathcal{G}_g $ and $\bar{\eta}_{b,g,k}(z,x)=\bar{m}_{b,j,k}(z.x)/a_{b,j}^2,\,\forall j\in \mathcal{G}_g$ that allows the asymptotic expressions for the elements of the coupling matrix and consequently the asymptotic ICI expressions to be simplified. In particular, 
the derivative of $\bar{\eta}_{b,g,k}(z,x)$ with respect to $x$ at point $x=0$ and $z=-1$ can be evaluated similar to Appendix~\ref{sec:proof Theorem2} from 
 \begin{equation}
 \label{eq:eta proof deriv1}
 \begin{aligned}
\bar{\eta}_{b,g,k}^{\prime}=
\,\,\,\,\frac{1}{N}{ \rm Tr} \bigg({\boldsymbol{\Theta}_{b,g} \mathbf{T}_{b,g}}
\big( \sum_{j \in \mathcal{G}_g} \!\!\frac{(\bar{\lambda}_j a_{b,j}^2)^2\boldsymbol{\Theta}_{b,g}    \bar{\eta}_{b,g,k}^{\prime}}{N(1+\bar{\lambda}_j a_{b,j}^2 \bar{\eta}_{b,g})^2} +a_{b,k}^2\boldsymbol{\Theta}_{b,g} \big) \mathbf{T}_{b,g}\bigg)
\end{aligned}
\end{equation}
with
 \begin{equation}
{\mathbf{T}_{b,g}}={\bigg(\frac{1}{N} \sum_{j \in \mathcal{G}_g} \frac{\bar{\lambda}_j a_{b,j}^2 \boldsymbol{\Theta}_{b,g}}{1+
		\bar{ \lambda}_ja_{b,j}^2\bar{\eta}_{b,g}
	}+\mathbf{I}_{N} \bigg)^{-1}}
\end{equation}
where then similar to~\eqref{eq:numer G}-\eqref{eq:Tbkl prime}, the unknown variable $\bar{\eta}_{b,g,k}'$ can be solved as
\begin{equation}\label{eq:eta prime 2idx}
	\bar{\eta}_{b,g,k}^{\prime}=\frac{a_{b,k}^2}{N}\frac{{\rm Tr}  ((\boldsymbol{\Theta}_{b,g} \mathbf{T}_{b,g})^2)}{1-\rho_{b,g} \rm Tr  ((\boldsymbol{\Theta}_{b,g} \mathbf{T}_{b,g})^2)}
\end{equation}
with $\rho_{b,g}=\frac{1}{N^2} \sum_{j \in \mathcal{G}_g} ({\bar{\lambda}_j a_{b,j}^2   })^2/{(1+\bar{\lambda}_j a_{b,j}^2 \bar{\eta}_{b,g})^2}$.
 Similarly, the normalization terms $\|{\bf v}_j\|^2=\frac{1}{N^2}{\vec h}_{b,j}\herm \boldsymbol{\Sigma}^{\backslash j}_{b} \boldsymbol{\Sigma}^{\backslash j}_{b} {\vec h}_{b,j},\forall j \in \mathcal U_{b}\cap \mathcal{G}_g$ converges almost surely as 
 \begin{equation}
 \label{eq:norm V}
 \|{\bf v}_j\|^2-\frac{a_{b,j}^2}{N} \bar \zeta_{b,g}' \rightarrow 0
 \end{equation}
 where the measure ${\zeta}_{b,g}(x)$ is given as ${\zeta}_{b,g}(x)=\frac{1}{N} \text{Tr} (\boldsymbol{\Theta}_{b,g} ( \sum_{j\in \mathcal U}\frac{\bar{\lambda}_{j}}{N}\mathbf{  h}_{b,j}\mathbf{  h}_{b,j}\herm  +(1-x)\mathbf{I}_N)^{-1}) $, and similar to~\eqref{eq:eta proof deriv1}-\eqref{eq:eta prime 2idx}, the deterministic equivalent for derivative of ${\zeta}_{b,g}(x)$ with respect to $x$ at $x=0$ can be evaluated as
 \begin{equation}\label{eq:eta prime}
\bar{\zeta}_{b,g}^{\prime}=\frac{{\rm Tr}  \big(({\boldsymbol{\Theta}}_{b,g} {\mathbf{T}}_{b,g})^2\big)/N}{1-\rho_{b,g} {\rm Tr}  ((\boldsymbol{\Theta}_{b,g} \mathbf{T}_{b,g})^2)}.
\end{equation}
Thus, the deterministic equivalent for ${\epsilon_{b,k}}$ can be evaluated, based on~\eqref{eq:epsil group exact} and~\eqref{eq:norm V}, as $\bar \epsilon_{b,k}=-\sum_{j\in \mathcal U_{b}\cap \mathcal{G}_g } \bar p_{j} N[\bar{\mathbf{G}}]_{k,j}/a_{b_j,j}^2 \bar \zeta_{b,g}'$. Recalling the non-diagonal elements of $\bar{\mathbf{G}}$ from~\eqref{eq:nonnormal G} and using $\bar{\eta}_{b,g,k}'=\bar{m}_{b,j,k}'/a_{b,j}^2,\forall j\in \mathcal{G}_g$, we get
\begin{equation}
\label{eq:ICI appen}
\bar \epsilon_{b,k}
=  \frac{(\bar{\eta}_{b,g,k}^{\prime})/(\bar{\zeta}_{b,g}^{\prime})}{(1+\bar{\lambda}_k a_{b,k}^2 \bar \eta_{b,g})^2}\sum_{j\in \mathcal U_{b}\cap \mathcal{G}_g }\!\! \bar p_{j} , \quad\forall k\in \mathcal{G}_g.
\end{equation}
Finally, replacing $\bar{\eta}_{b,g,k}^{\prime}$ and $\bar{\zeta}_{b,g}^{\prime}$ with equivalents in~\eqref{eq:eta prime 2idx} and~\eqref{eq:eta prime}, respectively, and denoting $P_{b,g}=\sum_{j\in \mathcal U_{b}\cap \mathcal{G}_g } \bar p_{j}$, the interference term $\bar \epsilon_{b,k}$ can be written as in the  corollary.

Next, we evaluate the total power required at a given BS to serve UEs within a group in asymptotic regime. In doing so, we start from the SINR constraints in~\eqref{Opt_problem}, which can be equivalently written for UE~$k$ as
\begin{equation}
\label{eq:norm SINR}
 \frac{p_k\left|{\bf h}_{b_k,k}\herm{\bf v}_{k}\right|^{2}/\|{\bf v}_k\|^2}{\epsilon_{b_k,k} + \sum_{b\in{\mathcal B}\setminus b_{k}} \epsilon_{b,k} +  \sigma^{2}}   \geq {\gamma_{k}}
\end{equation}
where $\epsilon_{b_k,k} $ denotes the intra-cell interference and $\epsilon_{b,k},\,\forall b\in{\mathcal B}\setminus b_{k} $ are the intercell interference term. Denoting the SINR of $k^\text{th}$ UE by $\Gamma_k$, we have $\Gamma_k-\bar{\Gamma}_k\rightarrow 0$ almost surely with
\begin{equation}
\bar{\Gamma}_k=\frac{	
	({N a_{b_k,k}^2 \bar{\eta}_{b_k,g_k}^2}/{\gamma_k}{\bar{\zeta}_{b_k,g_k}^{\prime}} )\bar{p}_k
	}{\bar{\epsilon}_{b_k,k} + \sum_{b\in{\mathcal B}\setminus b_{k}} \bar{\epsilon}_{b,k} +  \sigma^{2}}
\end{equation}
 where the deterministic equivalents for numerator and denominator of~\eqref{eq:norm SINR} directly follows from~\eqref{eq:nonnormal G} and~\eqref{eq:ICI appen}.
Since the SINR constraints at the optimal point must be satisfied with equality, we set $\bar{\Gamma}_k=\gamma_k$ to evaluate $\bar{p}_k$ as
\begin{equation}
\label{eq:pk group}
\bar{p}_k=
	\frac{\bar{\zeta}_{b_k,g_k}^{\prime}}{\bar{\eta}_{b_k,g_k}^2}
\frac{\gamma_k}{N a_{b_k,k}^2 } 
({\bar{\epsilon}_{b_k,k} + \sum\limits_{b\in{\mathcal B}\setminus b_{k}} \bar{\epsilon}_{b,k} +  \sigma^{2}}).
\end{equation}
Now, consider BS $b'$ with a subset of UEs within group~$g'\in \mathcal{A}_{b'}$. The transmit power imposed on BS $b'$ for serving UEs $k\in\mathcal{G}_{g'}\cap \mathcal{U}_{b'}$ is given by $\bar{P}_{b',g'}=\sum_{k\in \mathcal U_{b'}\cap \mathcal{G}_{g'}} \bar p_{k}$, with $\bar{p}_k$ given by~\eqref{eq:pk group}. Keeping this in mind and plugging~\eqref{eq:ICI appen} into~\eqref{eq:pk group}, we get a system of equation to evaluate $P_{b',g'}$ as follows
\begin{equation}
\label{eq:sys eq group P}
\begin{aligned}
\bar{P}_{{{b'}},{g'}}
\frac{\bar{\eta}_{{{b'}},{g'}}^2} {\bar{\zeta}_{{{b'}},{g'}}^{\prime}}
=  \frac{1}{N}\sum_{k\in \mathcal U_{{{b'}}}\cap \mathcal{G}_{g'}} \frac{\gamma_k}{ a_{{{b'}},k}^2 } \sigma^2+ 
\sum_{b\in \mathcal{B}}\sum_{g\in \mathcal{A}_b}\sum_{k\in \mathcal U_{{{b'}}}\cap \mathcal{G}_{g'}\cap\mathcal{G}_g} \!\!\!\!\!\!\!\!\!\!\!
\frac{ {\rm Tr}  \big((\boldsymbol{\Theta}_{b,g} \mathbf{T}_{b,g})^2\big)/{\rm Tr}  (\boldsymbol{\Theta}_{b,g} \mathbf{T}_{b,g}^2)}{N(1+\frac{\gamma_ka_{b,k}^2 \bar \eta_{b,g}}{a_{{{b'}},k}^2 \bar \eta_{{{b'}},{g'}}})^2} \frac{\gamma_k{a_{b,k}^2} }{{a_{{{b'}},k}^2} }\bar{P}_{b,g} \\
\end{aligned}
\end{equation}
evaluated $\forall {{b'}}\in \mathcal{B},\forall {g'}\in\mathcal{A}_{b'}$.
By rearranging the above equations in matrix form, the system of equations in the corollary is obtained.

\bibliographystyle{IEEEtran}

\vspace{-0.8cm}
\bibliography{Jourbib}

\end{document}